\documentclass[11pt]{article}
\newcommand{\be}{\begin{equation}}
\newcommand{\ee}{\end{equation}}
\newcommand{\bea}{\begin{eqnarray}}
\newcommand{\eea}{\end{eqnarray}}
\newcommand{\sn}{{\rm sn}}

\newcommand{\dn}{{\rm dn}}
\newcommand{\cn}{{\rm cn}}
\newcommand{\sech}{{\rm sech}}
\usepackage[dvips]{graphicx}
\usepackage{amsfonts}
\usepackage{amssymb}
\usepackage{amsmath}
\usepackage{subfigure}

\begin{document}

\vspace{.5in}
\begin{center}
{\LARGE{\bf Mapping Between Generalized Nonlinear Schr\"odinger Equations and
Neutral Scalar Field Theories and New Solutions of the Cubic-Quintic NLS
Equation}}
\end{center}

\vspace{.3in}
\begin{center}
{\LARGE{\bf Avinash Khare}} \\
{Institute of Physics, Bhubaneswar, Orissa 751005, India}
\end{center}

\begin{center}
{\LARGE{\bf Avadh Saxena}} \\
{Theoretical Division and Center for Nonlinear Studies, Los
Alamos National Laboratory, Los Alamos, NM 87545, USA}
\end{center}

\begin{center}
{\LARGE{\bf Kody J.H. Law}} \\
{Department of Mathematics and Statistics,
University of Massachusetts, Amherst, MA 01003, USA}
\end{center}

\vspace{.9in}
{\bf {Abstract:}}

We
highlight an interesting mapping between the moving breather solutions
of the generalized Nonlinear Schrodinger (NLS) equations and the static
solutions of neutral scalar field theories. Using this connection, we
then obtain several new moving breather solutions of the cubic-quintic NLS
equation both with and without uniform phase in space.  The stability of
some stationary solutions is investigated numerically and the results
confirmed via dynamical evolution.

\newpage

\section{Introduction}
\label{intro}
The nonlinear Schr\"odinger equation is one of the most celebrated integrable
nonlinear equation which has found application in several areas of physics
including the self-focusing of intense Laser beams, Langmuir waves in Plasma,
the collapse of the Bose-Einstein condensates (BECs) with attractive
interactions etc \cite{ss}. While most of the applications so far are those
of NLS with cubic nonlinearity, in recent years the cubic-quintic NLS (CQNLS)
has also found several applications. Some of these are in nonlinear optics
and BEC. For example, in nonlinear optics, it describes the propagation of
pulses in double-doped optical fibres \cite{ang}
and in Bragg-gratings \cite{malomed}; in BEC it models the
condensate with two- and three-body interactions \cite{agtf,zwpm}.
It is thus of considerable interest to obtain exact solutions of the CQNLS
and other generalized NLS equations.

In particular, a considerable amount of attention has been given in the
recent decades to localized solutions of various NLS, although these
solutions exist as the hyperbolic limit of appropriate general elliptic
function solutions.  This direction has been explored in the context of
an external periodic potential \cite{bronski} as well as weakly interacting 
solitary waves in generalized NLS equations \cite{zy}.  Herein we explore
exhaustively the analytic elliptic function solutions in the absence
of an external potential.

In this context it is worth noting the
connection between the
moving breather solutions of the generalized NLS equations and the (static)
solutions of neutral scalar field theories, known as Galilean invariance.
Consider the generalized NLS equation
\be\label{1}
iu_t+u_{zz}+f(|u|^2)u=0\,,
\ee
where $f(.)$ is a real valued algebraic function with $f(0)=0$. On using the
ansatz
\be\label{2}
u(z,t)= \phi(z-vt-z_0)\exp{\{(i/2)v(z-kt-\delta)\}}\,,
\ee
where $k=(v^2+4a)/2v$, it is easily shown that
$u$ satisfies Eq. (\ref{1}) provided $\phi(x)$
satisfies the equation

\be\label{3}
\phi_{xx}+f(\phi^2)\phi+a\phi=0\,,~~x=z-vt-z_0\,,
\ee

\noindent which
is the static field equation for a scalar field theory with
nonlinear interaction term $f(\phi^2)\phi$.
This is a travelling wave solution, and can also be posed in $x$ coordinates
by the transformation $\partial_z \rightarrow \partial_x$ and
$\partial_t \rightarrow -v \partial_x + \partial_t$, in which case,
considering $z_0=0, \delta=0$ without any loss of generality,
Eq. (\ref{2}) can be considered separable in this frame as
\be\label{3.5}
u(x,t)= \phi(x)\exp{\{(i/2)v x\}}\exp{\{(i/2)(v^2-4a)t\}} =
\tilde{\phi}(x)\exp{\{-i\tilde{a} t\}} \,.
\ee

Now this Galilean transformed solution is a stationary solution in the traveling frame.
Actually, we can slightly generalize the above ansatz, i.e. consider instead
\be\label{2a}
u(z,t)= \phi(z-vt-z_0)\exp{\{(i/2)(vz-vkt-\delta+\eta(x))\}}\,,~~x=z-vt-z_0\,.
\ee
On equating real and imaginary parts, it is easily shown that,
given $g$, if
\be\label{2b}
\eta(x)=g\int_{0}^{x} \frac{dx'}{\phi^2(x')}\,,
\ee
and
\be\label{3a}
\phi_{xx}+f(\phi^2)\phi+a\phi=\frac{g^2}{4\phi^3}\,,~~x=z-vt-z_0\,,
\ee
with $4a=v(2k-v)$ as above, then (\ref{3a}) solves (\ref{1}).
(Note in the traveling frame of $x$ there would also be an extra nonlinear
term due to the first derivative, for $v \neq 0$.)

It thus follows from here that knowing the various static solutions of
the scalar field theory (\ref{3}) or (\ref{3a}), one can immediately
write down the moving
breather solutions of the corresponding NLS model by using Eqs. (\ref{1}) and
(\ref{2}). Note that the breather solution of the corresponding NLS model will
be uniquely characterized by its five parameters: $v,a,g,\delta$ and $z_0$.
In particular it follows from here that in order to obtain the breather
solutions of the CQNLS equation, which is characterized by
\be\label{4}
f(|u|^2)=b|u|^2+c|u|^4\,,
\ee
one needs to obtain the static solutions of the $\phi^2$-$\phi^4$-$\phi^6$ field
theory characterized by
\be\label{5}
\phi_{xx}+a\phi+b\phi^3+c\phi^5=0\,,
\ee
in case $g=0$ while if $g \ne 0$, then one needs to obtain the solutions of
the equation
\be\label{5a}
\phi_{xx}+a\phi+b\phi^3+c\phi^5-\frac{g^2}{4\phi^3}=0\,.
\ee


Note that in the special case of $c=0$, one gets the well known
mapping between
the celebrated (cubic) NLS and $\phi^2$-$\phi^4$ field theory. If instead one
wants to obtain the breather solutions of the quadratic-cubic NLS
characterized by
\be\label{6}
f(|u|^2)=b |u|+ c |u|^2\,,
\ee
then one needs to obtain the static solutions of the $\phi^2$-$\phi^3$-$\phi^4$
field theory characterized by
\be\label{7}
\phi_{xx}+a\phi+b\phi^2+c\phi^3=0\,.
\ee

We have thus transformed the problem of finding the breather solutions of
the generalized
NLS models to that of finding static solutions of neutral scalar field
theories with power law potentials. There is however, one important
difference. In view
of the fact that the potential must be bounded from below, in neutral scalar
field theory models, normally one
takes the coefficient of the leading term in the potential to be positive, i.e.
in Eqs. (\ref{5}) and (\ref{7}), one takes the coefficient $c<0$. However, in
the context of the generalized NLS equations,
this coefficient need not necessarily
be negative. In fact usually it is taken to be positive, yielding bright
solitons while if it is negative, then one obtains dark solitons. It is easily
seen that taking the coefficient $c>0$ in Eqs. (\ref{5}) and (\ref{7}) is
equivalent to considering the solutions of the nonlinear oscillator problem
(rather than the field theory problem).

The purpose of this paper is to give exhaustive solutions to the CQNLS as
well as quadratic-cubic NLS equation as given by Eq. (\ref{7}). For that
purpose we make use of the known static solutions of the corresponding
$\phi^2$-$\phi^4$-$\phi^6$ as well as $\phi^2$-$\phi^3$-$\phi^4$ field theory 
as well as oscillator models and also
obtain few new solutions, which to the best of our knowledge, had not been
explicitly written down before in the literature. Further, we also obtain
static solutions of the field theory $\phi^2$-$\phi^4$-$\phi^6$-$\phi^{-2}$.
 In this context, it is worth
pointing out that recently we have obtained several static solutions of the
coupled $\phi^2$-$\phi^4$-$\phi^6$ \cite{ks1}, $\phi^2$-$\phi^3$-$\phi^4$ \cite{ks2}
as well as  $\phi^2$-$\phi^4$ \cite{ks3} models from where we can immediately
obtain the solutions in the decoupled case, for $c<0$. Extending these
ideas, we also obtain solutions of these models even in case $c>0$.

The paper is organized as follows.
In Sec. \ref{cqnls}
we obtain
static solutions of the
constant (or linear in the case $v \neq 0$) phase CQNLS problem
(i.e. when $\phi^{-2}$ term is
absent).  In section \ref{num}
we investigate the linear stability of some of these solutions
numerically.  In particular, we focus on the hyperbolic limit
in which our results here are consistent with the well-known
Vakhitov-Kolokolov criterion \cite{vk} as well as earlier
results for a more general non-linearity \cite{kivshar,weinstein}.  Dynamical
evolution confirms the stability results and a connection is observed
between the nature of the nonlinearity and the behavior of the
unstable evolution.  In particular, in the case $c>0$, when
$b \leq 0$ the solutions tend to blow-up with a self-similar behavior, 
while unstable solutions for which $b>0$ do not (perturbations are additive).
Then, in Sec. \ref{cqnls_nzp} we provide
solutions
for the $\phi^2-\phi^4-\phi^6-\phi^{-2}$ model as given by Eq. (\ref{5})
in case
each of $a,b,c$ can be either positive or negative.
In
Appendix \ref{A1} we provide static  solutions
for the $\phi^2-\phi^3-\phi^4$ model as given by Eq. (\ref{7}) in case
each of $a,b,c$ can be either positive or negative. These will be relevant
in the context of the breather solutions of quadratic-cubic NLS.
For completeness,
in Appendix \ref{A2} we also provide static solutions
for the $\phi^2-\phi^4$ model as given by Eq. (\ref{6}) in case
each of $a,c$ can be either positive or negative. These will be relevant
in the context of the breather solutions of NLS. Finally, in Sec.
\ref{conclusion} we
summarize the results and indicate possible relevance of these results.

\section{Solutions of $\phi^2$-$\phi^4$-$\phi^6$ and hence CQNLS Model}
\label{cqnls}

\subsection{Theoretical development}
\label{theory_cqnls}

Let us consider solutions of field Eq. (\ref{5}). As explained above, once
these solutions are obtained, then the solution of the CQNLS equation are
immediately obtained from here by using Eqs. (\ref{1}), (\ref{2}) and
(\ref{4}). We list below eight distinct periodic solutions to the field
Eq. (\ref{5}), i.e. the trivial phase case of $g=0$.
In each case, we also mention the values of the parameters
$a,b,c$, in particular, if they are positive or negative.

\subsubsection{Dark soliton families}
\label{cqnls_ds}

{\bf Solution I}

It is easily shown \cite{ks1} that
\be\label{2.1}
\phi=A\sqrt{1+\sn(Bx+x_0,m)}\,,
\ee
is an exact solution to the field Eq. (\ref{5}) provided
\be\label{2.2}
(5m-1) B^2=-4a\,,~~(5m-1)A^2=-\frac{8ma}{b}\,,~~3(5m-1)b^2=64mac\,.
\ee
Thus this solution is valid provided $b>0,c<0$ while $a>(<$ or $=)\,0$
depending on if $m< (>$ or $=)\, 1/5$. Here $\sn(x,m)$ and $\cn(x,m)$, $\dn(x,m)$ 
denote Jacobi elliptic functions with modulus $m$ \cite{gr,bf}. 

In the limit $m=1$ the periodic solution (\ref{2.1}) goes over to
 the dark soliton solution
\be\label{2.3}
\phi=A\sqrt{1+\tanh(Bx+x_0)}\,,
\ee
provided
\be\label{2.4}
B^2=-a\,,~~A^2=-(2a/b)\,,~~b^2=(16/3)ac\,.
\ee
Thus the dark soliton solution exists to field Eq. (\ref{5}) provided
$a<0,b>0,c<0$.

{\bf Solution II}

It is easily shown \cite{ks1} that
\be\label{2.17}
\phi=\frac{A\sn(Bx+x_0,m)}{\sqrt{1-D\sn^2(Bx+x_0,m)}}\,,
\ee
is an exact solution to the field Eq. (\ref{5}) provided
\bea\label{2.18}
&&[3D-(1+m)] B^2=-a\,,~~bA^2=2[2D(1+m)-m-3D^2]B^2\,, \nonumber \\
&&\frac{3b^2}{4ac}=\frac{[2D(1+m)-m-3D^2]^2}{D(1-D)(m-D)[3D-(1+m)]}\,.
\eea
There are different constraints depending on the value of $D$. For example if
$D<0$, then this solution is valid provided $a>0,b<0,c>0$. On the other hand,
if $D>0$ then while the solution is only valid if $c<0$, the signs of $a,b$
depend on the value of $D$. For example, while $a< (>$ or $=)\, 0$ depending on
if $D > (<$ or $=)\, (1+m)/3$, $b < (>$ or $=)\, 0$ depending on if
$D < (>$ or $=)\, \frac{1+m-\sqrt{1-m+m^2}}{3}$. Summarizing
\bea\label{2.19}
&&0<D<\frac{1+m-\sqrt{1-m+m^2}}{3}\,,~~~a>0,b<0,c<0\,, \nonumber \\
&&\frac{1+m-\sqrt{1-m+m^2}}{3} < D < \frac{1+m}{3}\,,
~~~a>0,b>0,c<0\,, \nonumber \\
&&\frac{1+m}{3}<D<m\,,~~~a<0,b>0,c<0\,, \nonumber \\
&&D<0\,,~~~a>0,b<0,c>0\,.
\eea

In the limit $m=1$ the periodic solution (\ref{2.17}) goes over to
 the dark soliton solution
\be\label{2.20}
\phi=\frac{A\tanh(Bx+x_0)}{\sqrt{1-D\tanh^2(Bx+x_0)}}\,,
\ee
provided
\be\label{2.21}
(3D-2) B^2=-a\,,~~bA^2=2(3D-1)(1-D)B^2\,,~~
\frac{3b^2}{4ac}=\frac{(3D-1)^2}{D(3D-2)}\,.
\ee
Thus the dark soliton solution exists to field Eq. (\ref{5}) provided
the following constraints are satisfied
\bea\label{2.22}
&&0<D<\frac{1}{3}\,,~~~a>0,b<0,c<0\,, \nonumber \\
&&\frac{1}{3} < D < \frac{2}{3}\,,
~~~a>0,b>0,c<0\,, \nonumber \\
&&\frac{2}{3}<D<1\,,~~~a<0,b>0,c<0\,, \nonumber \\
&&D<0\,,~~~a>0,b<0,c>0\,.
\eea

{\bf Solution III}

It is easily shown \cite{ks1} that
\be\label{2.30}
\phi=\frac{A}{\sqrt{1-D\sn^2(Bx+x_0,m)}}\,,
\ee
is an exact solution to the field Eq. (\ref{5}) provided
\bea\label{2.31}
&&[3m-(1+m)D] B^2=-aD\,,~~DbA^2=2[D^2-2D(1+m)+3m]B^2\,, \nonumber \\
&&\frac{3b^2}{4ac}=\frac{[D^2-2D(1+m)+3m]^2}{(1-D)(m-D)[3m-(1+m)D]}\,.
\eea
There are different constraints depending on the value of $D$. For example if
$D<0$, then this solution is valid provided $a>0,b<0,c>0$. On the other hand,
if $D>0$ then while the solution is only valid if $c<0$, the signs of $a,b$
depend on the value of $D$. For example, while $a<0$ in case $m \ge 1/2$,
for $m<1/2$, $a< (>$ or $=)\, 0$ depending on
if $D < (>$ or $=)\, \frac{3m}{(1+m)}$. On the other hand,
$b > (<$ or $=)\, 0$ depending on if
$D < (>$ or $=)\, 1+m-\sqrt{1-m+m^2}$.

In the limit $m=1$ the periodic solution (\ref{2.30}) goes over to
 the dark soliton solution
\be\label{2.32}
\phi=\frac{A}{\sqrt{1-D\tanh^2(Bx+x_0)}}\,,
\ee
provided
\be\label{2.33}
(3-2D) B^2=-aD\,,~~DbA^2=2(3-D)(1-D)B^2\,,~~
\frac{3b^2}{4ac}=\frac{(3-D)^2}{(3-2D)}\,.
\ee
There are different constraints depending on the value of $D$. For example if
$D<0$, then this solution is valid provided $a>0,b<0,c>0$. On the other hand,
if $D>0$ then the solution is only valid if $a<0,b>0,c<0$.

It is worth pointing out that for $D<0$, solution (\ref{2.30}) is not an
independent solution but rather can be easily derived from the solution
(\ref{2.15}) below by using well known transformation formulas for Jacobi elliptic
functions and making use of the translational invariance of the solutions.
However, for $D>0$, (\ref{2.30}) is an independent solution.

\subsubsection{Bright soliton families}
\label{cqnls_bs}

{\bf Solution I}

It is easily shown that
\be\label{2.5}
\phi=A\sqrt{1+\cn(Bx+x_0,m)}\,,
\ee
is an exact solution to the field Eq. (\ref{5}) provided
\be\label{2.6}
(4m+1) B^2=4a\,,~~(4m+1)A^2=-\frac{8ma}{b}\,,~~3(4m+1)b^2=64mac\,.
\ee
Thus this solution is valid provided $a>0,b<0,c>0$.

In the limit $m=1$ the periodic solution (\ref{2.5}) goes over to
 the bright soliton solution
\be\label{2.7}
\phi=A\sqrt{1+\sech(Bx+x_0)}\,,
\ee
provided
\be\label{2.8}
5B^2=4a\,,~~5A^2=-(8a/b)\,,~~15b^2=64ac\,.
\ee
Thus the bright soliton solution exists to field Eq. (\ref{5}) provided
$a>0,b<0,c>0$.

{\bf Solution II.1}

It is easily shown that
\be\label{2.9}
\phi=A\sqrt{1+\dn(Bx+x_0,m)}\,,
\ee
is an exact solution to the field Eq. (\ref{5}) provided
\be\label{2.10}
(4+m) B^2=4a\,,~~(4+m)A^2=-\frac{8a}{b}\,,~~3(4+m)b^2=64ac\,.
\ee
Thus this solution is valid provided $a>0,b<0,c>0$.

In the limit $m=1$ the periodic solution (\ref{2.9}) goes over to
 the bright soliton solution (\ref{2.7}) satisfying the constraints
(\ref{2.8}).

{\bf Solution II.2}

It is easily shown that
\be\label{2.11}
\phi=A\sqrt{\dn(Bx+x_0,m)+k'}\,,~~k'=\sqrt{1-m}\,,
\ee
is an exact solution to the field Eq. (\ref{5}) provided
\be\label{2.12}
(4-5m) B^2=4a\,,~~bA^2=-2B^2k'\,,~~cA^4=(3/4)B^2\,.
\ee
Thus this solution is valid provided $b<0,c>0$ while $a > (<$ or $=)\,0 $
depending on if $m < (>$ or $=)\, (4/5)$. .

In the limit $m=1$ the periodic solution (\ref{2.11}) goes over to
 the bright soliton solution
\be\label{2.13}
\phi=A\sqrt{\sech(Bx+x_0)}\,,
\ee
provided
\be\label{2.14}
B^2=-4a\,,~~b=0\,,~~cA^4=(3/4)B^2\,.
\ee
Thus the bright soliton solution exists to field Eq. (\ref{5}) provided
$a<0,b=0,c>0$.

{\bf Solution II.3}

It is easily shown that
\be\label{2.15}
\phi=A\sqrt{\dn(Bx+x_0,m)-k'}\,,~~k'=\sqrt{1-m}\,,
\ee
is an exact solution to the field Eq. (\ref{5}) provided
\be\label{2.16}
(4-5m) B^2=4a\,,~~bA^2=2B^2k'\,,~~cA^4=(3/4)B^2\,.
\ee
Thus this solution is valid provided $b>0,c>0$ while $a > (<$ or $=)\,0$
depending on if $m < (>$ or $=)\, (4/5)$. .

In the limit $m=1$ the periodic solution (\ref{2.15}) goes over to
 the bright soliton solution (\ref{2.13}) satisfying the constraints
(\ref{2.14}).

{\bf Solution III}

It is easily shown \cite{ks1} that
\be\label{2.23}
\phi=\frac{A\cn(Bx+x_0,m)}{\sqrt{1-D\sn^2(Bx+x_0,m)}}\,,
\ee
is an exact solution to the field Eq. (\ref{5}) provided
\bea\label{2.24}
&&[2m-1-D(2-m)] B^2=-(1-D)a\,,~~(1-D)bA^2=2[(1+2D)m-D(D+2)]B^2\,, \nonumber \\
&&\frac{3b^2}{4ac}=\frac{[(1+2D)m-D(D+2)]^2}{D(m-D)[2m-1-D(2-m)]}\,.
\eea
There are different constraints depending on the value of $D$. For example if
$D<0$, then this solution is valid provided $c>0$. On the other hand,
the signs of $a,b$
depend on the value of $D<0$. For example, while for $m\ge 1/2$, $a<0$, for
$m<1/2$, $a > (<$ or $=)\, 0$ depending on if
$|D| < (>$ or $=)\, \frac{1-2m}{2-m}$.
On the other hand $b > (<$ or $=)\, 0$ provided
$|D| < (>$ or $=)\, 1-m +\sqrt{1-m+m^2}$.

On the other hand, if $0 < D < m$, then this solution is valid if $c<0$,
while the signs of $a,b$ depend on the value of $D>0$. In particular,
while $a>0$ in case $m \le 1/2$, for $m>1/2$, $a<0$ in case
$0<D<\frac{2m-1}{2-m}$ while $a>0$ provided $\frac{2m-1}{2-m} < D < m$.
On the other hand,  $b>0$ if
$0<D < -(1+m)+\sqrt{1-m+m^2}$, $b<0$ in case
$-(1+m)+\sqrt{1-m+m^2} < D < m$.

In the limit $m=1$ the periodic solution (\ref{2.23}) goes over to
 the bright soliton solution
\be\label{2.25}
\phi=\frac{A\sech(Bx+x_0)}{\sqrt{1-D\tanh^2(Bx+x_0)}}\,,
\ee
provided
\be\label{2.26}
B^2=-a\,,~~bA^2=2(1+D)B^2\,,~~
\frac{3b^2}{4ac}=\frac{(1+D)^2}{D}\,.
\ee
Thus the bright soliton solution exists to field Eq. (\ref{5}) provided
the following constraints are satisfied
\bea\label{2.27}
&&0<D<1\,,~~~a<0,b>0,c<0\,, \nonumber \\
&&-1 < D < 0\,,~~~a<0,b>0,c>0\,, \nonumber \\
&&D<-1\,,~~~a<0,b<0,c>0\,.
\eea

{\bf Solution IV}

It is easily shown \cite{ks1} that
\be\label{2.28}
\phi=\frac{A\dn(Bx+x_0,m)}{\sqrt{1-D\sn^2(Bx+x_0,m)}}\,,
\ee
is an exact solution to the field Eq. (\ref{5}) provided
\bea\label{2.29}
&&[2m-m^2-D(2m-1)] B^2=-(m-D)a\,, \nonumber \\
&&(m-D)bA^2=2[(1-2D)m-D(D-2)]B^2\,, \nonumber \\
&&\frac{3b^2}{4ac}=\frac{[(1-2D)m-D(D-2)]^2}{D(1-D)[2m-m^2-D(2m-1)]}\,.
\eea
There are different constraints depending on the value of $D$. For example if
$0<D<m$, then this solution is valid provided $a<0,b>0,c<0$.

On the other hand, if $ D < 0$, then this solution is valid if $c>0$,
while the signs of $a,b$ depend on the value of $D<0$. In particular,
while $a<0$ in case $m \ge 1/2$, for $m<1/2$, $a<0$ in case
$0<|D|<\frac{m(2-m)}{1-2m}$ while $a>0$ provided $\frac{m(2-m)}{1-2m} < D < m$.
On the other hand,  $b>0$ if
$0<D < -(1-m)+\sqrt{1-m+m^2}$ while $b<0$ in case
$-(1-m)+\sqrt{1-m+m^2} < D < m$.

In the limit $m=1$ the periodic solution (\ref{2.28}) goes over to
 the bright soliton solution (\ref{2.25}) satisfying the constraints
as given by Eqs. (\ref{2.26}) and (\ref{2.27}).

It is worth pointing out that for $D>0$, solution (\ref{2.28}) is not an
independent solution but rather can be easily derived from the solution
(\ref{2.23}) by using well known transformation formulas for Jacobi elliptic
functions and making use of the translational invariance of the solutions.
However, for $D<0$, (\ref{2.28}) is an independent solution.

\subsection{Numerical Solutions and Stability Analysis}
\label{num}

We will now look at some of the particular solutions presented above and
investigate their stability properties both through a systematic linear
stability analysis and also dynamical evolution.  In particular, we
consider exhaustively the various regimes of hyperbolic dark and
bright soliton solutions with trivial phase from section \ref{cqnls},
as well as some elliptic examples from the same section.
In considering the structural linearized stability we will
consider the static reference frame (i.e. v=0) and
focus on the stationary wave solutions. Such stationary
solutions have Galilean invariance, yielding the ansatz for the traveling wave given in Eq. (\ref{3}).  Although,
assuming a traveling reference frame and
Galilean transformations of the perturbations as well,
the equations for dynamical stability of static ($v=0$)
and traveling wave ($v \neq 0$) solutions are equivalent.
We note that above the Landau critical velocity, the
solution will be so-called Landau-unstable \cite{landau}, although this
instability is not expected to have any effect in the absence of any external potential, and is usually
investigated in the presence of an external impurity
\cite{sykes} or lattice \cite{wu}.Given a solution $\phi$ of (\ref{3}),
we consider the linearization of Eq. (\ref{1}) around this stationary solution, i.e. assume the
following ansatz, for $\epsilon \ll 1$,

\begin{figure}
\includegraphics[width=100mm]{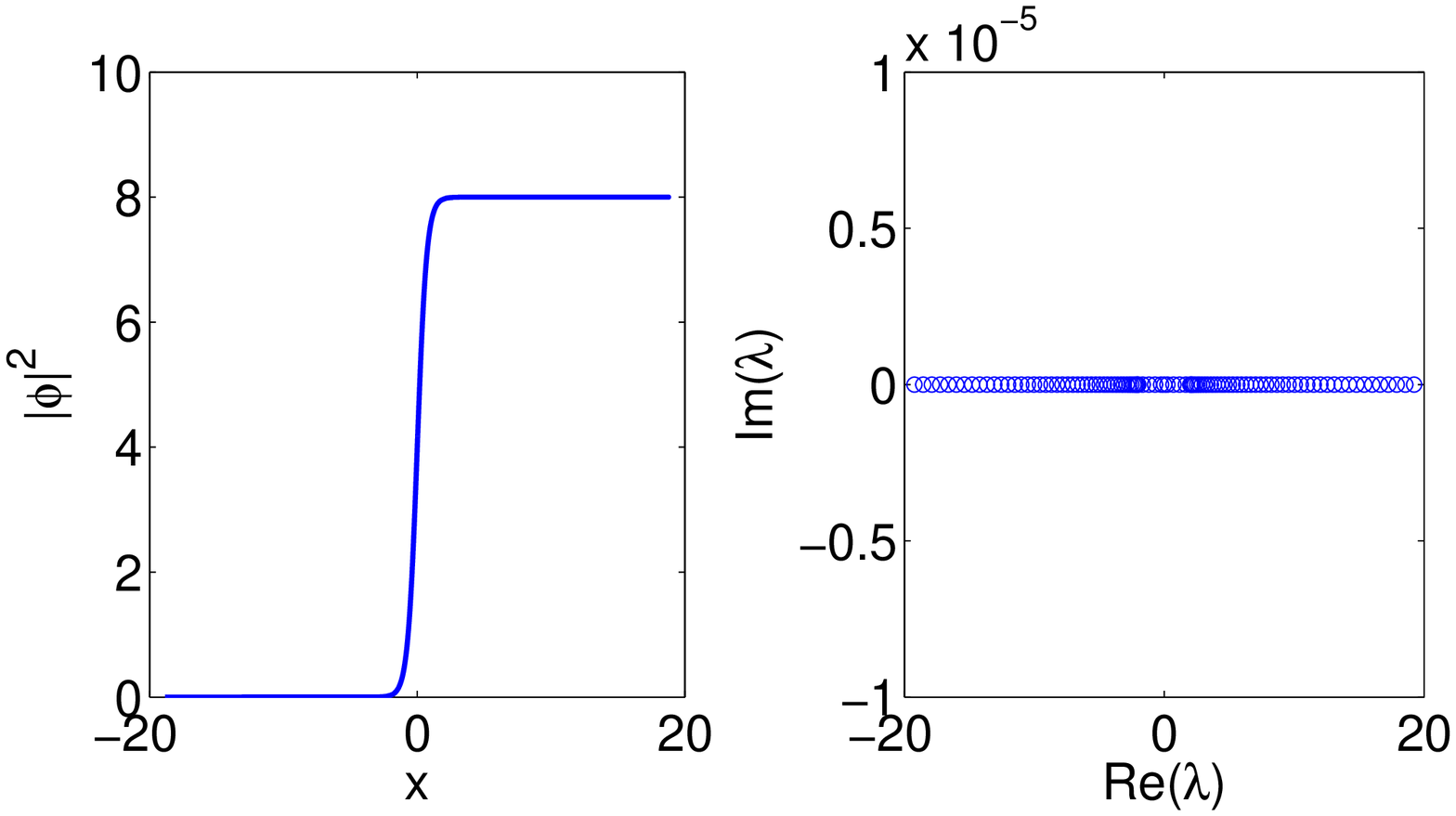}
\includegraphics[width=70mm]{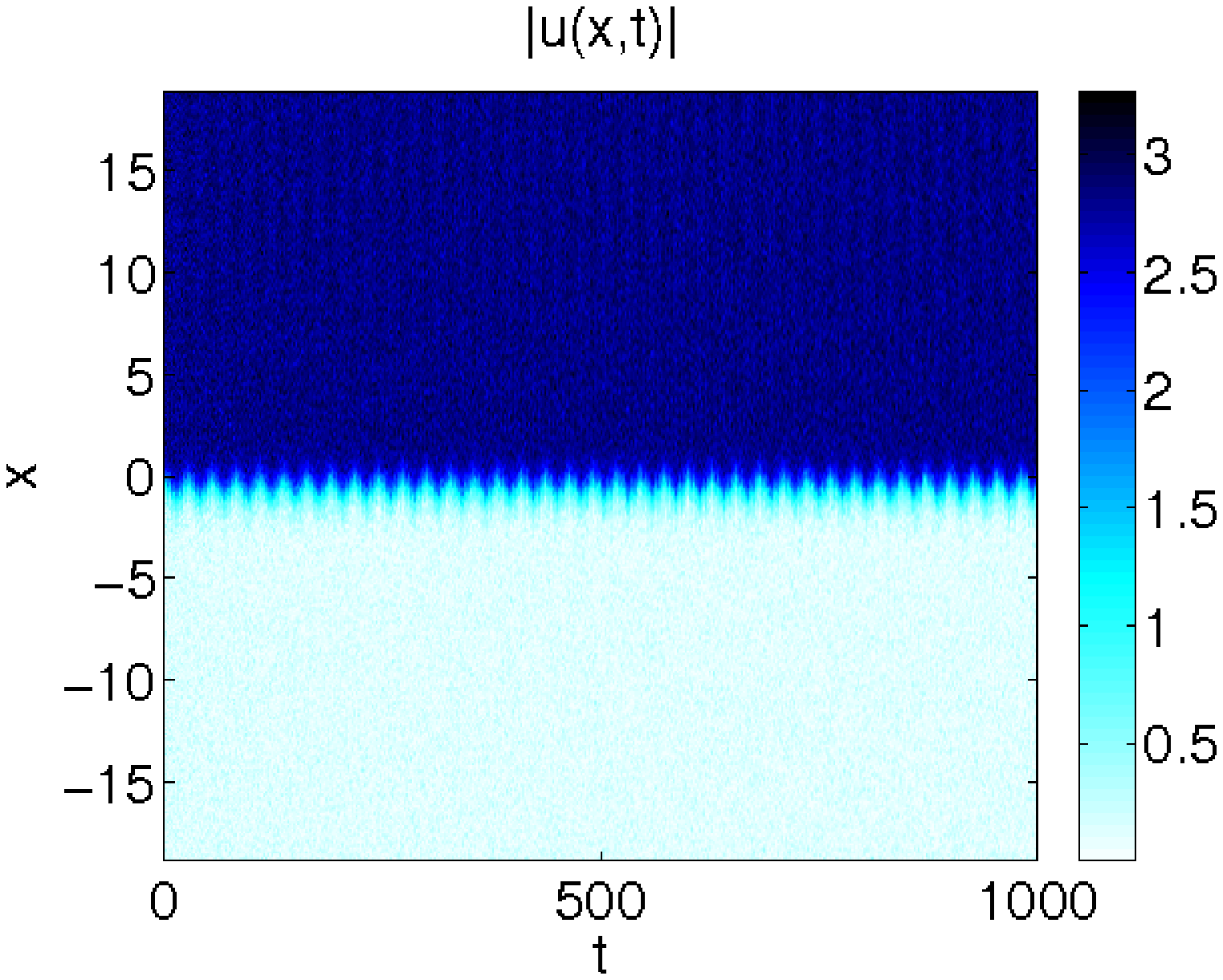}
\caption{(Color Online) Left: Dark soliton solution I with
$m=1,a=-2,b=1$ (left), and
the linearization spectrum as computed by Eq. (\ref{2.2.1}).
Right: Dynamical evolution of
$u=\phi+r$, where $r=p {\rm max} |\phi(x,0)|^2 U$, $p=0.05$,
$U$ is a uniformly distributed random variable in $(0,1)$,
and $\phi$ is solution I with $m=1,a=-2,b=1$, as shown
to the left.
The stability is confirmed as the solution sustains only minor
oscillations from its original form after the $5\%$ perturbation. }
\label{F1}
\end{figure}

\be\label{2.2.1}
u(z,t)=e^{-iat}[\phi(z)+\epsilon \tilde{u}(z,t)]\,.
\ee

Assuming $\tilde{u}=w(x) e^{i \lambda t}$ is separable then we
have the following linearization system

\begin{center}
\begin{math}\label{2.2.2}
\bordermatrix{& \cr &\mathcal{L}_1 &\mathcal{L}_2 \cr
 &-\mathcal{L}_2^* &-\mathcal{L}_1^* \cr}
v =
\lambda v
\end{math}
\end{center}
for any arbitrary eigenfunction $v$ (which includes the possibility of a  more 
general spatial form of the perturbations). The blocks are given by
\begin{eqnarray*}
\mathcal{L}_1 &=& a+\partial_{xx} + 2b|\phi|^2 + 3c|\phi|^4 \, , \\
\mathcal{L}_2 &=& b \phi^2 + 2c |\phi|^2 \phi^2.
\end{eqnarray*}

This eigenvalue problem, commonly known as the
Bogoliubov system, has been solved with both the
matlab functions
{\bf eig} in which the full matrix is diagonalized and
{\bf eigs}, which
implements the standard ARPACK Arnoldi iterative algorithm to solve
for the smallest/largest $n$ eigenvalues.
In particular, the latter, quicker and more efficient
method for evaluating a particular subset of eigenvalues,
is used for exhaustive continuations and benchmarking
over much finer spatial grids,
while agreement is found in all cases in which both
methods are used.

\subsubsection{Dark soliton families}
\label{ds_num}

For the first family of solutions given by Eqs. (\ref{2.1},\ref{2.3}),
the hyperbolic tangent [$m=1$, Eq. (\ref{2.3})]
(also known as ``kink'', or ``dark soliton'') solution was found to be stable,
just like in the case of the cubic NLS.  An example solution
is presented in the left panels of Fig. \ref{F1} for $a=-2$ and $b=1$.
Continuations
of these solutions for $(a,b) \in [-2,-.1] \times [.01, 1]$
were performed and there was no deviation from this stability result.
It seems reasonable
to conclude that this is universal, although we did not do an
exhaustive search of parameter space.  The stability results were
then confirmed with dynamical evolution using a standard fourth
order Runge-Kutta method.  The right panel of
Figure (\ref{F1}) displays the result of
evolving $u=\phi+r$, where $r=p {\rm max} |\phi(x,0)|^2 U$, $p=0.05$
and $U$ is a uniformly distributed random variable in $(-1,1)$.
Notice even with this large perturbation, the robust structure persists
for up to at least $t=1000$, experiencing only minor oscillations around
the stationary solution.


It is most likely a result of the fact that the solutions for $m \neq 1$
are ``very nearly non-differentiable" that, even with appropriately matched
periodic boundary conditions, the norm difference of the exact solution
with the one on the numerical grid is orders of magnitude larger than
that of the hyperbolic kin.  This prevents such a systematic
numerical analysis of the stability, although using the exact solution
as an initial guess for a fixed point solver on the numerical grid,
the solution does converge
to a smoother version which is wildly unstable, and this suggests
these solutions are unstable.

\begin{figure}
\includegraphics[width=80mm]{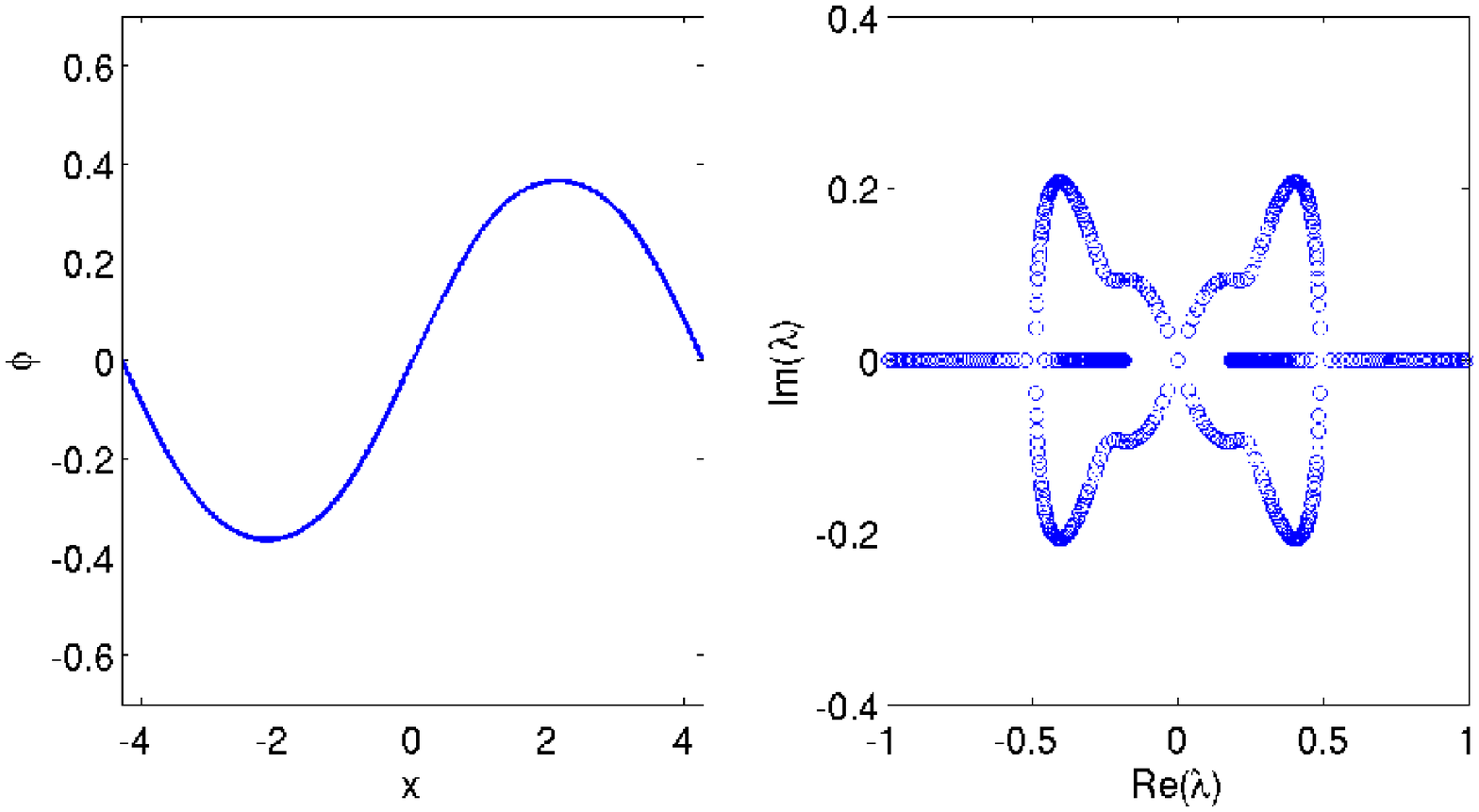}
\includegraphics[width=90mm]{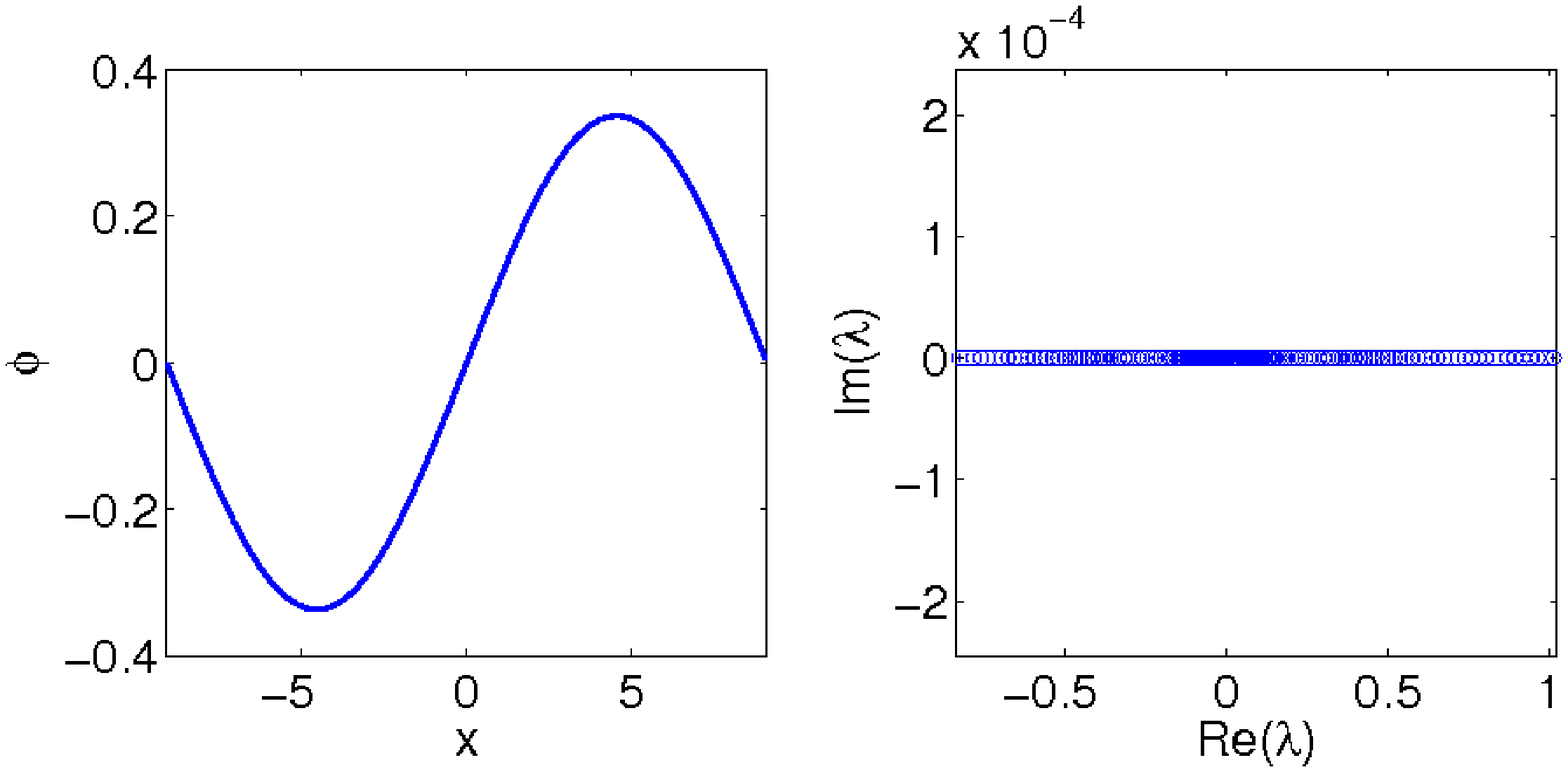}\\
\caption{(Color Online) The left two panels show the dark soliton solution II
(left) with unstable parameters
$(a,b,D,m)=(0.75,1,0.3,0.7)$
and the linearization spectrum
is on the right as given by Eqs. (\ref{2.2.1}).  The right
panels show a solution from the same family but for a set of
parameters for which
the solution is stable, $(a,b,D,m)=(0.1,0.5,0.3,0.5)$.}
\label{F4_2}
\end{figure}

\begin{figure}
\includegraphics[width=90mm]{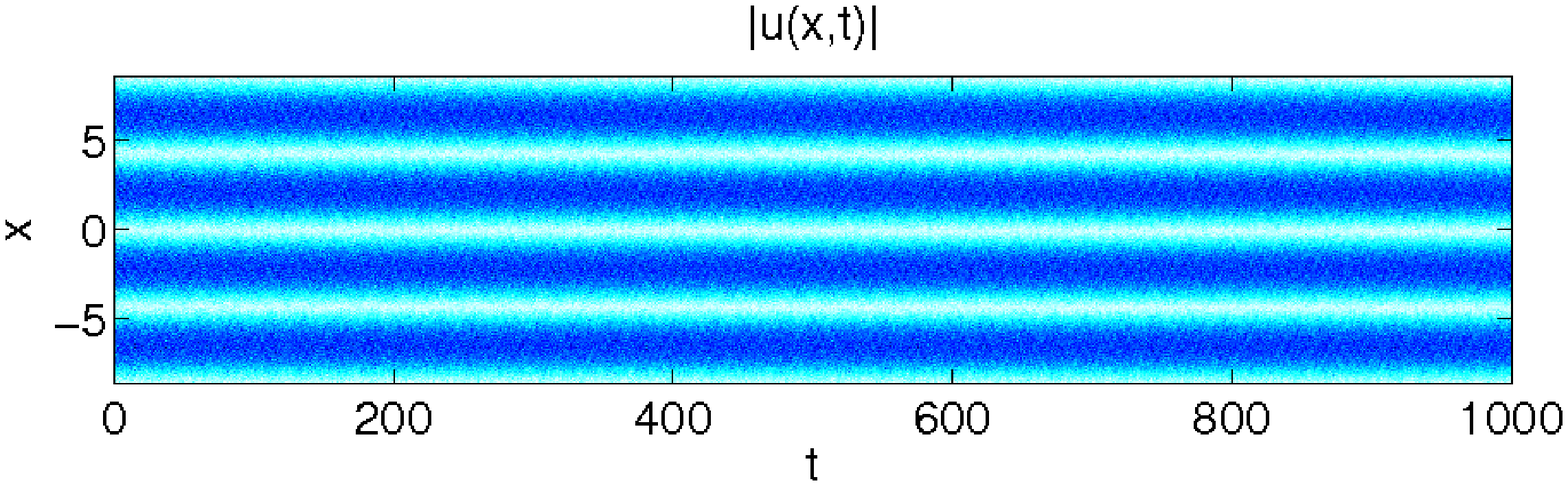}
\includegraphics[width=90mm]{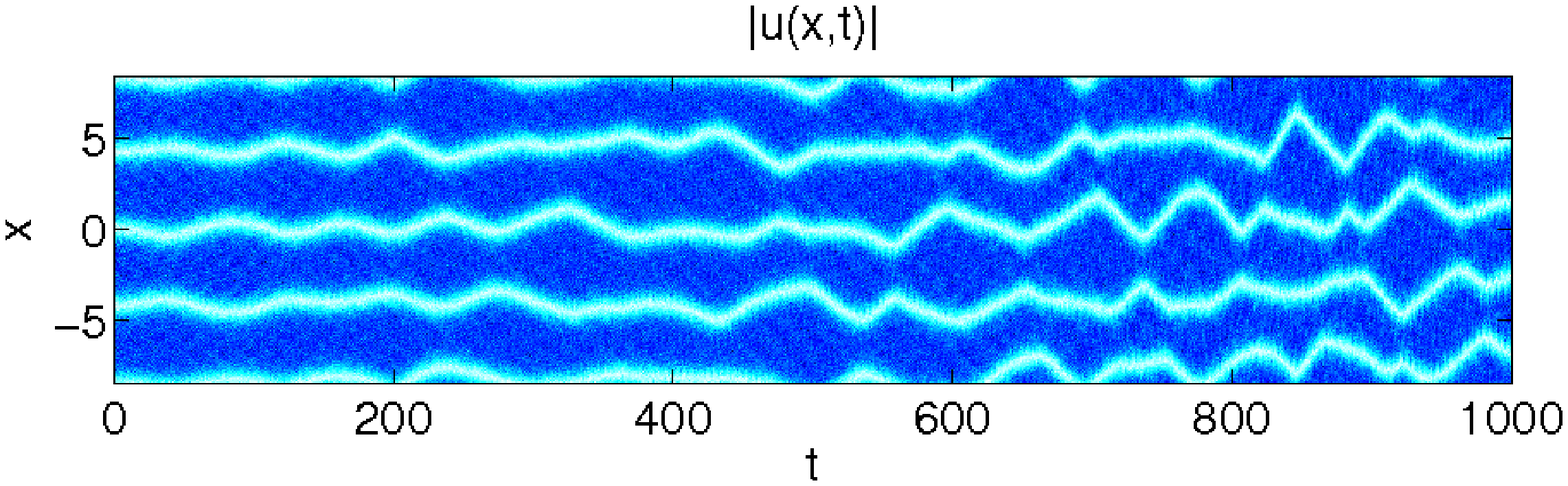}
\caption{(Color Online) The left panel is the dark soliton solution II
with unstable parameters $(a,b,D,m)=(0.75,1,0.3,0.7)$ from the left
of Fig. \ref{F4_2}, while the right is for the unstable
parameters $(a,b,D,m)=(1.5,1.5,0.5,0.99)$.}
\label{F4_3}
\end{figure}

Next, we consider the smoother family of dark-soliton solutions
II given in Eqs.
(\ref{2.17},\ref{2.20}).  For the hyperbolic solutions (m=1), Eq.
(\ref{2.20}), a representative value of $D$ was chosen
for all the intervals defined in Eq. (\ref{2.22}), with the exception
of the interval $2/3<D<1$, where the signs of $b$ and $c$ are the
same as solution I above. For each of these values of
$D \in \{ -0.2,0.1,0.35\}$, a two parameter
continuation was done for the corresponding values of
$(|a|,|b|) \in [.1,1] \times [.01,1]$.  All solutions were
found to be stable.
Some periodic solutions ($m\ne1$) from this smoother family
were also found.  For certain parameters, the solutions
are actually stable, while for others, they are very
unstable.
The unstable solution for $(a,b,D,m)=(0.75,1,0.3,0.7)$,
and the stable one for $(a,b,D,m)=(0.1,0.5,0.3,0.5)$
as given in Eq. (\ref{2.17}), and their linearization
spectra are shown in Fig. (\ref{F4_2}).
We note here that
the Floquet theorem has been invoked in conjunction with periodic
boundary conditions, since the stability matrix of the periodic solution
is periodic, in order to scan the infinite energy spectrum with a
subdivision of $200$ separate spatial frequencies out of the infinitely many.
Notice the continuum of eigenvalues with negative imaginary part which form
loops symmetric with respect to both the real and imaginary axes (the symmetry
is a result of the fact that the matrix is Hamiltonian).
The dynamics of the unstable solution with $(a,b,D,m)=(0.75,1,0.3,0.7)$
from the left panels of Fig. \ref{F4_2} are shown on the left panel of
Fig. \ref{F4_3}.  The linear instability has almost no discernible effect
on the evolution even with a very large $25\%$ perturbation of the initial
amplitude.  Somehow the stability properties of the hyperbolic limit
are ``inherited'' by its family, regardless of stability of the linearized
system.  This presumably has to do with this solution residing
at a small steep local maximum of the energy within a large basin,
in which a small perturbation is enough to leave the linear regime,
but not to significantly alter the structure of the solution (the unstable
eigenfunctions are presumably of a very similar form as the solution).
We believe this ``non-linear stability" is an interesting phenomenon,
but is outside the scope of this paper and will be investigated further
elsewhere.
The right panel shows the evolution of another unstable solution
from this family for parameter values $(a,b,D,m)=(1.5,1.5,0.5,0.99)$,
where the instability
has a greater effect, but still rather negligible and at
long times considering the large
perturbation.  Many other unstable solutions from this family were evolved
with large perturbations and this was the most unstable among them.


\subsubsection{Bright soliton families}
\label{bs_num}

\begin{figure}
\includegraphics[width=80mm]{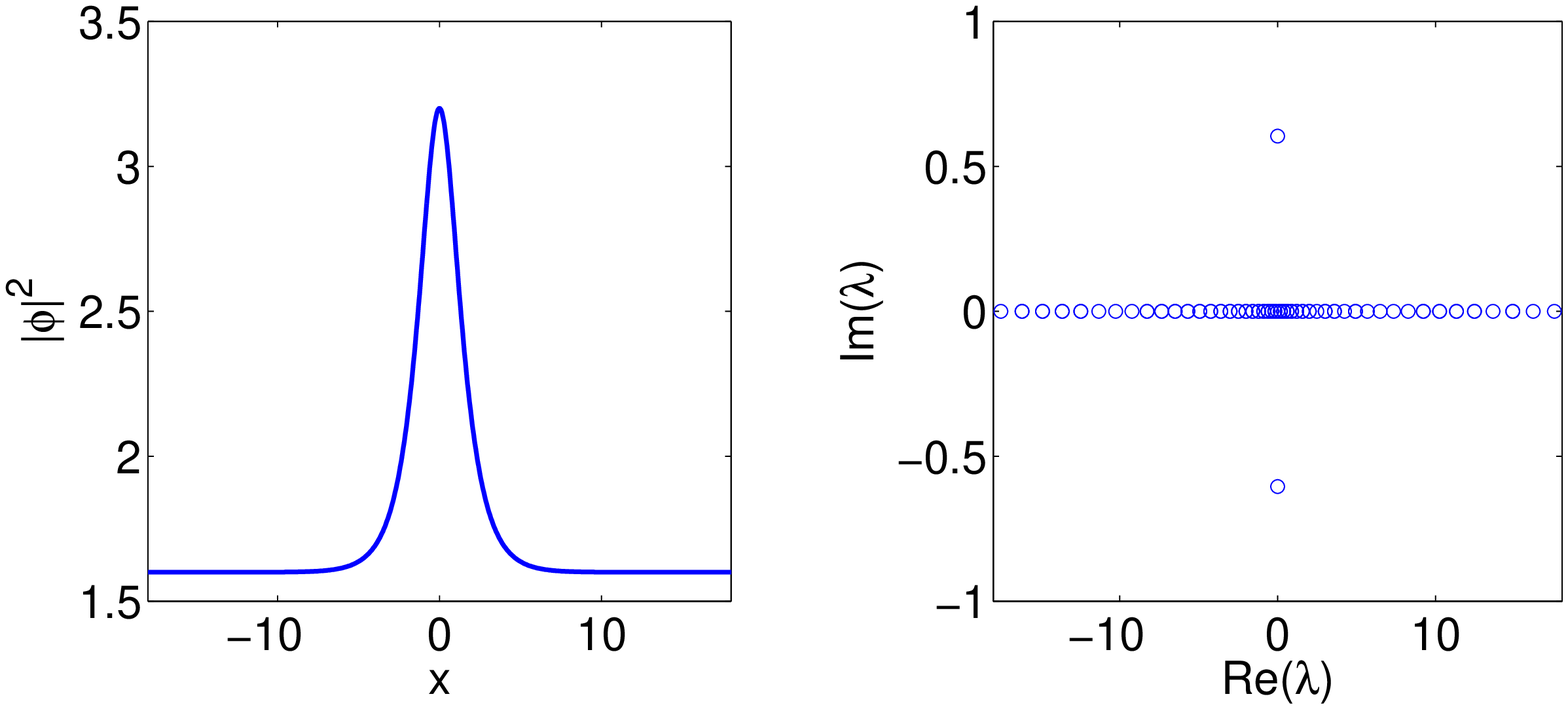}
\includegraphics[width=100mm]{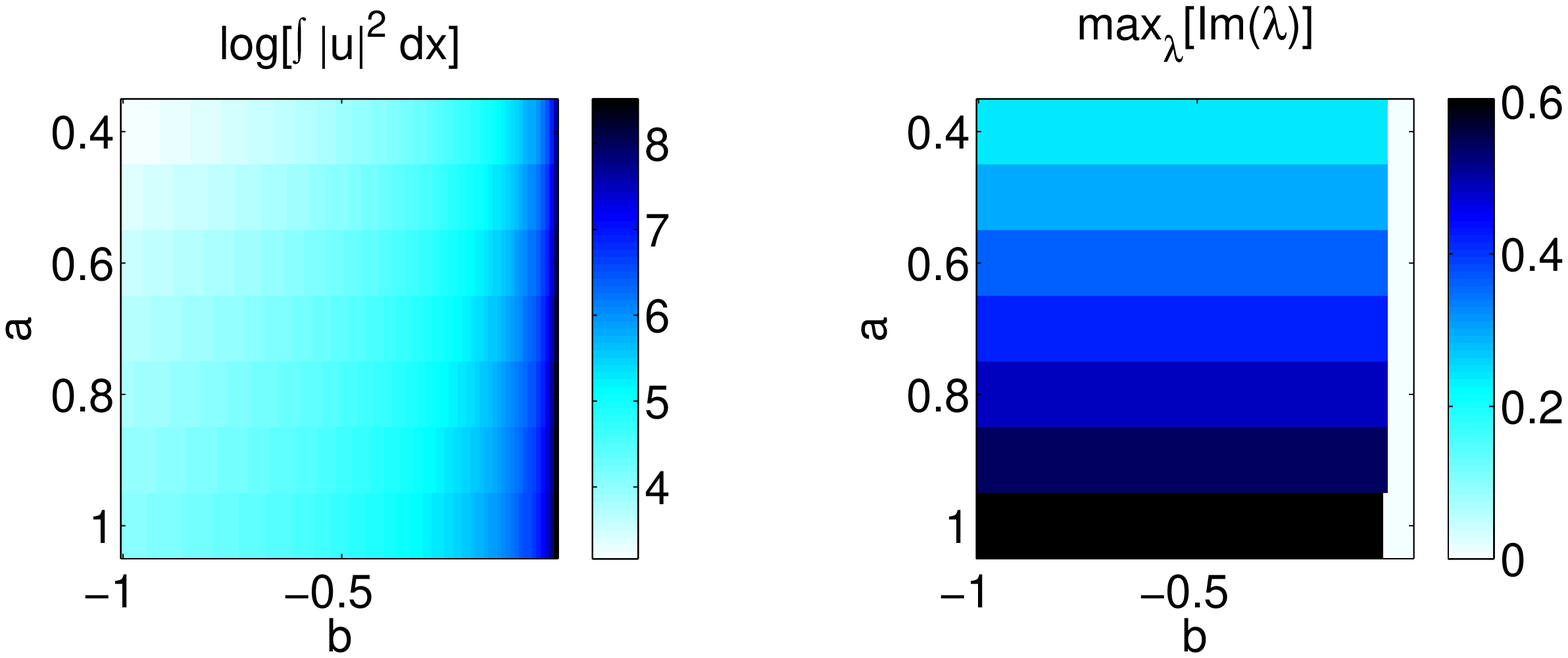}
\caption{(Color Online) Left panels are the bright soliton
solution I with $m=1,a=1,b=-1$
(left), and its linearization spectrum.  Right panels depict
the power (left, given on a log scale)
and stability (right), as indicated
by the maximum imaginary part of the spectrum, of a two-parameter
family of bright-soliton solutions I (of Eq. (29) with $m=1$ and
$(a,b) \in [0.4,1] \times [-0.01,-1]$.}
\label{F5}
\end{figure}


\begin{figure}
\includegraphics[width=100mm]{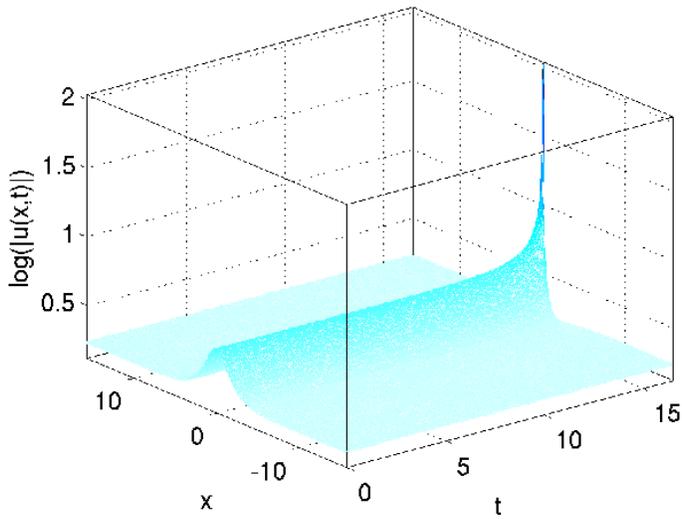}
\caption{(Color Online) Dynamical evolution of
$u=\phi+r$, where $r=p {\rm max} |\phi(x,0)|^2 U$, $p=0.05$, $U$ is a uniformly distributed random variable in $(0,1)$
and $\phi$ is solution I with $m=1,a=1,b=-1$, as
shown if Fig. (\ref{F5}).
The stability is confirmed and the solution collapses with
exponentially self-similar nature.
A $\log$ scale is used for the intensity to illustrate this.}
\label{F7}
\end{figure}

Next we turn to the bright soliton solutions.
Now the hyperbolic [$m=1$, Eq. \ref{2.7}] solution
(namely ``pulse'', ``bright soliton'', or just ``soliton'')
is actually unstable (see, e.g. \cite{kivshar}).
This is markedly different from the cubic NLS case, where the analogous solution
is stable.  This can be connected to the fact that it is well-known
that the $H^1$ norm (or, energy) of the solution to Eq. (\ref{1})
for $f(x)=x^{2 \sigma}$ is bounded for $\sigma<2/d$, where $d$ is the
dimensionality of the problem \cite{weinstein}.  The value $\sigma$ is known as the
{\it critical exponent}, and in our case of $d=1$,
we have $\sigma=4$.  This is the smallest
power nonlinearity for which blowup can occur, corresponding to an
exact balance between kinetic and potential energies under the
constraint of conserved mass.
The solution and its stability are presented in
the left panels of Fig. (\ref{F5}).
A two-parameter continuation in
the parameters $a$ and $b$ (with $c$ determined by them) was performed
in order to confirm that the family of solutions is always unstable
in the region $(a,b) \in [0.4,1] \times [-1,-0.01]$,
as presented in the right panels of
Fig. (\ref{F5}).
The stability here can also be understood in terms of the
Vakhitov-Kolokolov criterion \cite{vk}, since $dN/da>0$.
The small region of stable solutions with very large amplitudes for
very small $b$ are actually constant solutions, since
for increasing $b$ the amplitude of the soliton shrinks,
while the ``pedestal" (plane wave) it is sitting on grows.
The dynamical evolution given in Fig.(\ref{F7})
confirms the prediction
and the solution collapses around $t=15$.
The collapse of the waveform is
presented on a $\log$ scale so it can be appreciated.

\begin{figure}
\includegraphics[width=100mm]{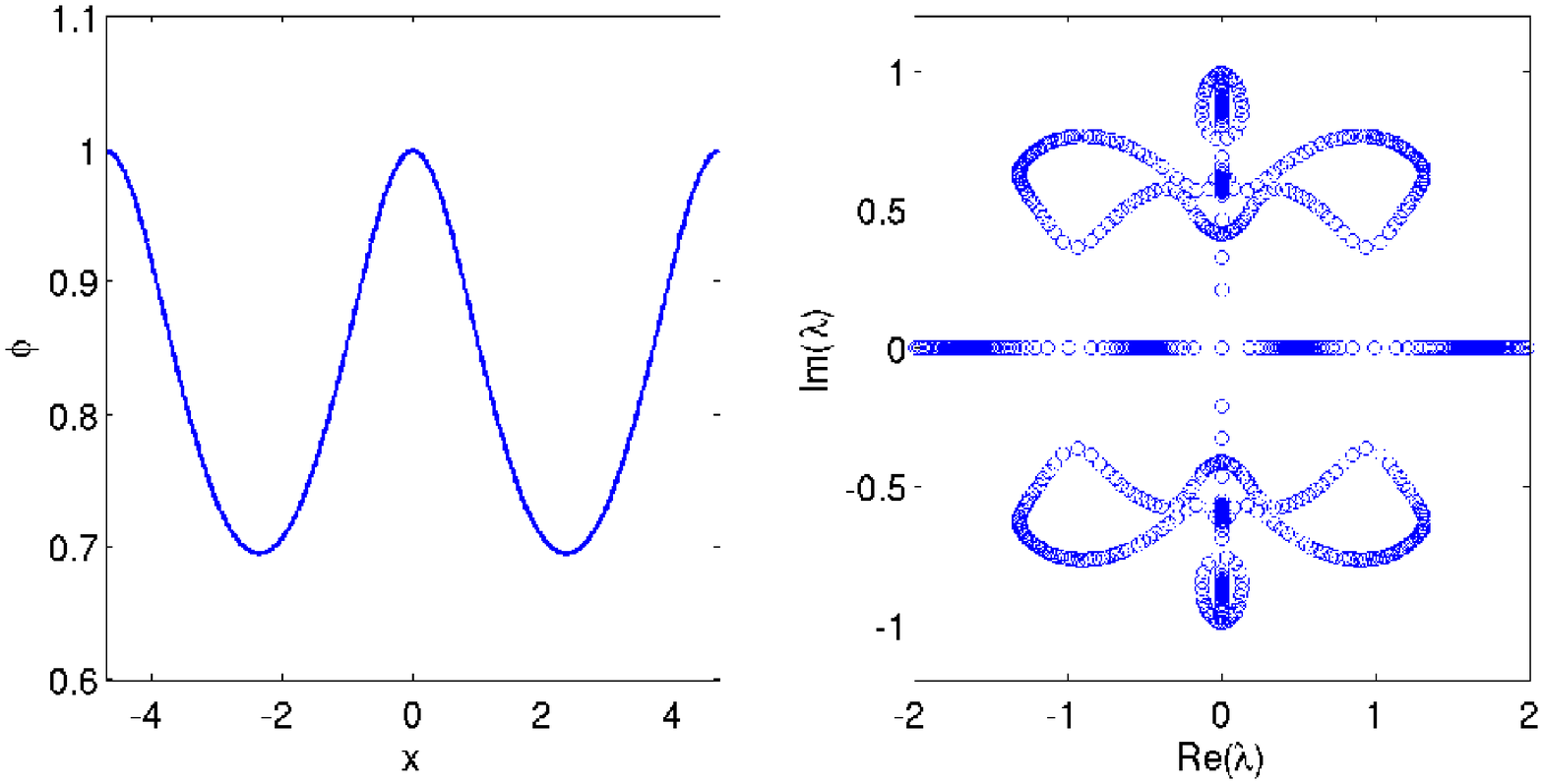}
\includegraphics[width=70mm]{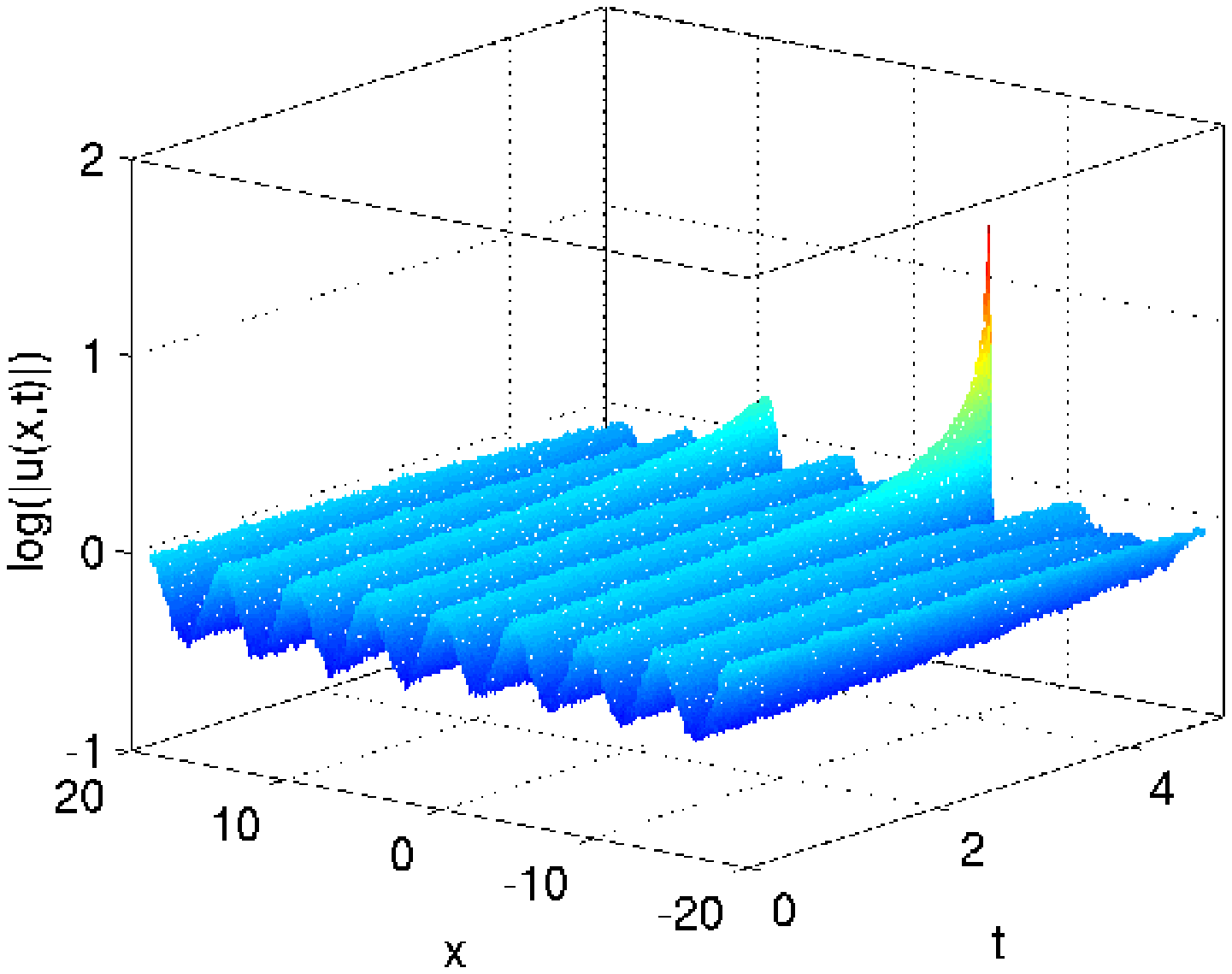}\\
\includegraphics[width=100mm]{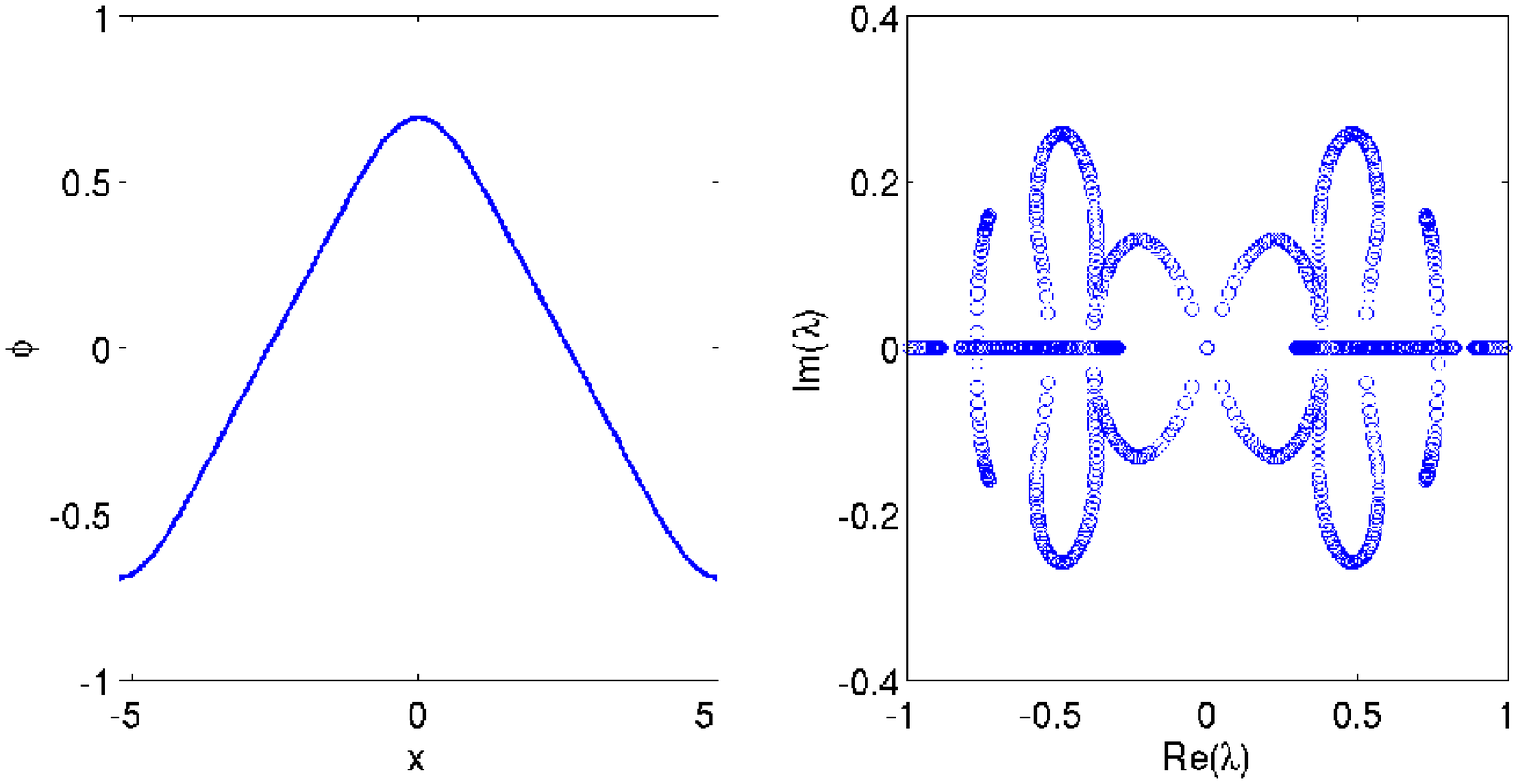}
\includegraphics[width=70mm]{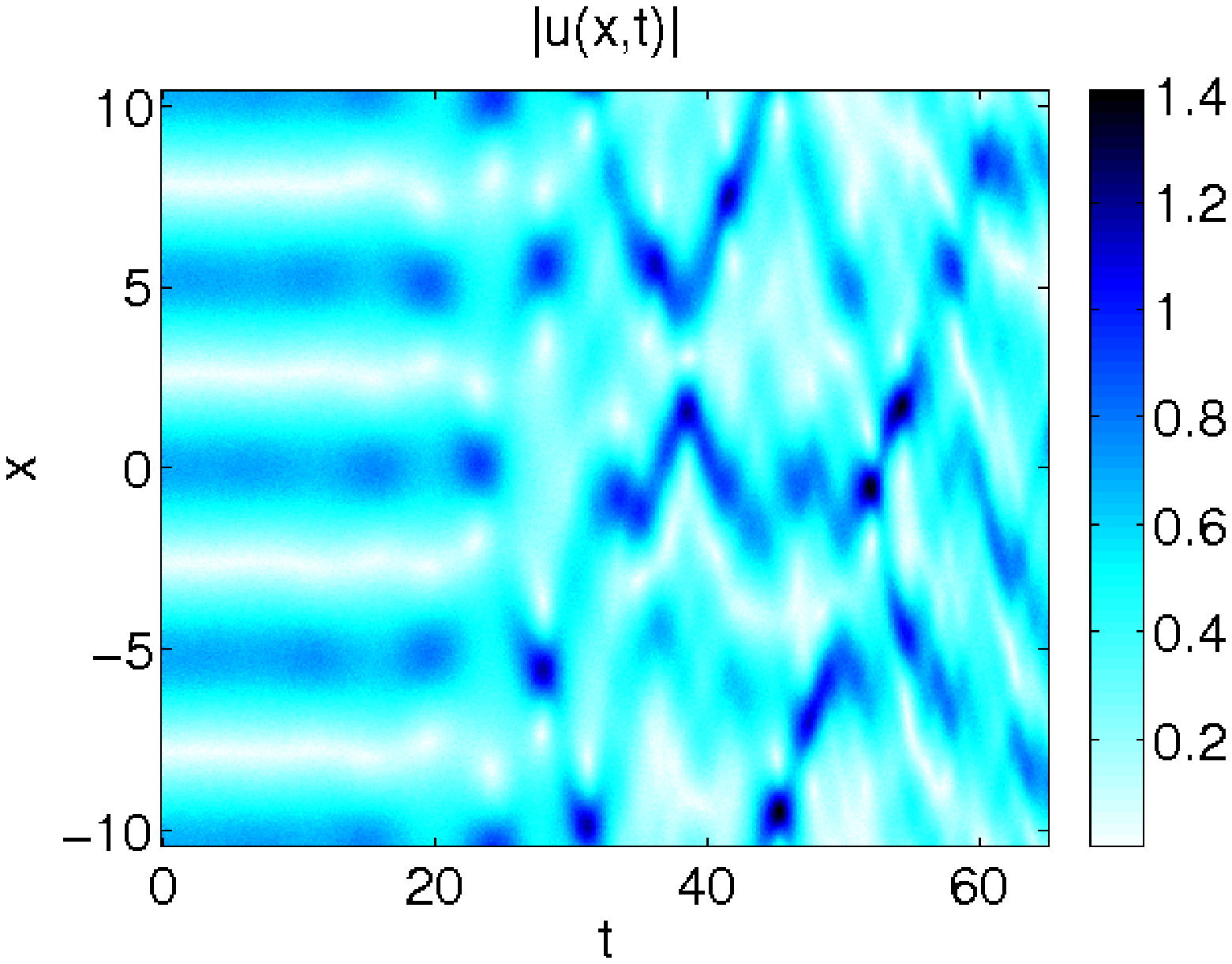}
\caption{(Color Online) Top: bright soliton solution II.2 with
$(a,b,m)=(-0.15,-1,0.9)$, ($c \approx 1.56$).  The linearization spectrum
is given in the middle as defined in Eqs. (\ref{2.2.1}), and
the dynamical evolution is to the right.
Bottom: Bright soliton solution III with $(a,b,m,D)=(-0.1,1,0.4,-0.3)$,
($c \approx 0.98$).}
\label{F8}
\end{figure}


It is interesting to note that the instability to collapse
appears to be correlated to the sign of the cubic term for the
bright solitons (again consistent with the findings of
\cite{kivshar}).  That is
to say, those bright solitons we found for which $b>0$ were
stable, while for $b<0$ they were unstable
($c>0$
 of course).
In particular, we now turn to solutions of Eq.
(\ref{2.23}).  
Two parameter continuations were performed for
$D\in\{-1.5,-0.3,-0.1,0.1,0.6\}$ and $(|a|,|b|)\in[0.1,1]\times[0.01,1]$,
and the last four were invariably stable, for all of which $b>0$
(for the former two $c>0$ and for the latter two $c<0$).  For the first
value of $D$ the corresponding $b$ is negative (and of course $c>0$) and,
as in solution I, the solution is unstable in the entire parameter region.
It should be noted that the solution in Eq. (\ref{2.13}) where
$b=0$ and $c>0$ is linearly stable as the imaginary pair of eigenvalues
for $b<0$ pass through the origin and become real for $b<0$.
This extra pair of eigenvalues is in addition to one pair
associated to phase invariance, $\delta$ in Eq. (\ref{2}), and one pair 
associated to translational invariance, $z_0$ in Eq. (\ref{2}),
and is due to the conformal invariance
of the soliton solution of the critical NLS.
This solution is, hence, unstable to collapse
as well, despite its linear stability.

Most solutions were found to be unstable, and in fact, no bright soliton 
family solutions were
found that are stable, although in the high-dimensional parameter
space among the infinitely many solutions, we by no means discount the
possibility that such solutions may exist.
As examples we present
the solution II.2 for the values $(a,b,m)=(-0.15,-1,0.9)$ as given
in Eq. (\ref{2.11})
and the solution III for values $(a,b,m,D)=(-0.1,1,0.4,-0.3)$.
The solution
with $b<0$ [solution II.2, Fig. (\ref{F8})]
does exhibit collapse phenomenon like
the unstable soliton, while the one with
$b>0$ [solution III] does not.


\section{Solutions of $\phi^2-\phi^4-\phi^6-\phi^{-2}$ and hence CQNLS Model}
\label{cqnls_nzp}

Let us consider solutions of field Eq. (\ref{5a}). As explained above, once
these solutions are obtained, then the solution of the CQNLS equation are
immediately obtained from here by using Eqs. (\ref{1}), (\ref{2a}),
(\ref{2b}), (\ref{3a}) and
(\ref{4}). We list below five distinct solutions to the field
Eq. (\ref{5a}) out of which four are periodic.

{\bf Solution I}

It is easily shown that
\be\label{8.1}
\phi=\sqrt{\frac{A\sn(Gx+x_0,m)+B}{D\sn(Gx+x_0,m)+F}}\,,
\ee
is an exact solution to the field Eq. (\ref{5a}) provided the following five
coupled equations are satisfied:
\be\label{8.2}
mG^2(AF-BD)(3AF+BD)=g^2 D^4-4A^2(aD^2+bAD+cA^2)\,,
\ee
\bea\label{8.3}
&&G^2(AF-BD)[2mBF+(1+m)AD]=2g^2 D^3 F-4aAD(AF+BD) \nonumber \\
&&-2bA^2(AF+3BD)-8cA^3 B\,,
\eea
\bea\label{8.4}
&&(1+m)G^2(AF-BD)^2=-6g^2 D^2 F^2+4a(A^2 F^2+B^2 D^2+4ABDF) \nonumber \\
&&+12bAB(AF+BD)+24cA^2 B^2\,,
\eea
\bea\label{8.5}
&&G^2(AF-BD)[2AD+(1+m)BF]=-2g^2 DF^3+4aBF(AF+BD) \nonumber \\
&&+2bB^2(3AF+BD)+8cAB^3\,,
\eea
\be\label{8.6}
-G^2(AF-BD)(AF+3BD)=g^2F^4-4B^2(aF^2+bBF+cB^2)\,.
\ee
Thus we have five coupled equations relating the five parameters $A,B,D,G,g$
(note that we can always remove one of the parameter from the ansatz
(\ref{8.1})). In the special case of $m=1$, the solution (\ref{8.1}) goes
over to the hyperbolic soliton solution
\be\label{8.7}
\phi=\sqrt{\frac{A\tanh (Gx+x_0)+B}{D\tanh (Gx+x_0)+F}}\,,
\ee
while in the limit $m=0$ it goes over to the trigonometric solution
\be\label{8.8}
\phi=\sqrt{\frac{A\sin (Gx+x_0)+B}{D\sin (Gx+x_0)+F}}\,.
\ee

In the special case of $D=0$, $F=1$, we obtain a simpler solution
\be\label{8.9}
\phi=\sqrt{A\sn(Gx+x_0,m)+B}\,,
\ee
provided
\be\label{8.10}
a<0, ~c<0\,,~~G^2=\frac{9b^2-32|a||c|}{8|c|(1+m)}\,,~~A^2=\frac{3mG^2}{4|c|}\,,
~~B=\frac{3b}{8|c|}\,,
\ee
while $g$ for this solution is given by
\bea\label{8.11}
&&g^2=4B^2(a+bB+cB^2)-A^2 G^2 \nonumber \\
&&=\frac{3}{[32(1+m)]^2 |c|^3}[192b^2 |a||c|(10m-1-m^2) \nonumber \\
&&-(64|a||c|)^2 m -9b^4(26m-5-5m^2)]\,.
\eea
It is easily checked that so long as $g \ne 0$, such a solution does not
exist for $m=1$, even though it exists for a range of values of $m<1$.
Also note that the solution (\ref{8.9}) does not exist in the
trigonometric limit of $m=0$ unless $c=0$.

{\bf Solution II}

Yet another solution to the field Eq. (\ref{5a}) is
\be\label{8.12}
\phi=\sqrt{\frac{A\cn(Gx+x_0,m)+B}{D\cn(Gx+x_0,m)+F}}\,,
\ee
provided the following five coupled equations are satisfied:
\be\label{8.13}
mG^2(AF-BD)(3AF+BD)=-g^2 D^4+4A^2(aD^2+bAD+cA^2)\,,
\ee
\bea\label{8.14}
&&G^2(AF-BD)[2mBF+(2m-1)AD]=-2g^2 D^3 F \nonumber \\
&&+4aAD(AF+BD) +2bA^2(AF+3BD)+8cA^3 B\,,
\eea
\bea\label{8.15}
&&(2m-1)G^2(AF-BD)^2=6g^2 D^2 F^2-4a(A^2 F^2+B^2 D^2+4ABDF) \nonumber \\
&&-12bAB(AF+BD)-24cA^2 B^2\,,
\eea
\bea\label{8.16}
&&G^2(AF-BD)[2AD(1-m)-(2m-1)BF]=-2g^2 DF^3 \nonumber \\
&&+4aBD(AF+BD) +2bB^2(3AF+BD)+8cAB^3\,,
\eea
\be\label{8.17}
-(1-m)G^2(AF-BD)(AF+3BD)=g^2F^4-4B^2(aF^2+bBF+cB^2)\,.
\ee
In the special case of $m=1$, the solution (\ref{8.12}) goes
over to the hyperbolic soliton solution
\be\label{8.18}
\phi=\sqrt{\frac{A\sech (Gx+x_0)+B}{D\sech (Gx+x_0)+F}}\,,
\ee
while in the limit $m=0$ it goes over to the trigonometric solution
\be\label{8.19}
\phi=\sqrt{\frac{A\cos (Gx+x_0)+B}{D\cos (Gx+x_0)+F}}\,.
\ee

In the special case of $D=0$, $F=1$, we obtain a simpler solution
\be\label{8.20}
\phi=\sqrt{A\cn(Gx+x_0,m)+B}\,,
\ee
provided
\be\label{8.21}
c>0\,,~~G^2=\frac{9b^2-32ac}{8c(2m-1)}\,,~~A^2=\frac{3mG^2}{4c}\,,
B=-\frac{3b}{8c}\,,
\ee
while $g$ for this solution is given by
\bea\label{8.22}
&&g^2=4B^2(a+bB+cB^2)-(1-m) A^2 G^2 \nonumber \\
&&=\frac{3}{[32(2m-1)]^2 c^3}[192b^2 ac(1+8m-8m^2) \nonumber \\
&&-(64ac)^2 m(1-m)-9b^4(5+16m-16m^2)]\,.
\eea
Thus unlike solution (\ref{8.1}), such a solution always exists
in the hyperbolic
limit so long as $32ac<9b^2<(192/5)ac$. Further, it also exists for a range
of values of $m$, though it does not exist in the $m=0$ limit, unless $c=0$.

In the special case of $m=1/2$, we obtain a one-parameter family of solutions
of the form (\ref{8.20}). In particular, in that case it follows from
Eq. (\ref{8.21}) that
\be\label{8.23}
9b^2=32ac\,,~~A^2=\frac{3G^2}{8c}\,,~~B=-\frac{3b}{8c}\,,~~
g^2=\frac{a^2}{3c}-\frac{4cA^4}{3}\,.
\ee
It is worth noting that while $A,G$ are arbitrary with their ratio being fixed,
the only constraint that
we have is that $A^2<\frac{a}{2c}$ so that $g$ is nonzero and real.

Unlike the $\sn$ case, in this case solutions exist even in case $A=0$
(and assuming $B=1$ without any loss of generality). In particular,
\be\label{8.24}
\phi=\frac{1}{\sqrt{D\cn(Gx+x_0,m)+F}}\,,
\ee
is an exact solution to Eq. (\ref{5a}) provided
\bea\label{8.25}
&&mG^2=g^2 D^2\,,~~3(1-m)D^2 G^2=g^2 F^4-4(aF^2+bF+c)\,, \nonumber \\
&&(2m-1)G^2=2(3g^2 F^2-2a)\,,~~4g^2 F^3=b+4aF\,.
\eea
Note that such a solution exists in the hyperbolic limit ($m=1$)
but not in the trigonometric limit ($m=0$).

{\bf Solution III}

Yet another solution to the field Eq. (\ref{5a}) is
\be\label{8.26}
\phi=\sqrt{\frac{A\dn(Gx+x_0,m)+B}{D\dn(Gx+x_0,m)+F}}\,,
\ee
provided the following five coupled equations are satisfied:
\be\label{8.27}
-G^2(AF-BD)(3AF+BD)=g^2 D^4-4A^2(aD^2+bAD+cA^2)\,,
\ee
\bea\label{8.28}
&&G^2(AF-BD)[2BF+(2-m)AD]=-2g^2 D^3 F+4aAD(AF+BD) \nonumber \\
&&+2bA^2(AF+3BD)+8cA^3 B\,,
\eea
\bea\label{8.29}
&&(2-m)G^2(AF-BD)^2=6g^2 D^2 F^2-4a(A^2 F^2+B^2 D^2+4ABDF) \nonumber \\
&&-12bAB(AF+BD)-24cA^2 B^2\,,
\eea
\bea\label{8.30}
&&G^2(AF-BD)[2AD(1-m)-(2-m)BF]=-2g^2 DF^3 \nonumber \\
&&+4aBD(AD+BF) +2bB^2(3AF+BD)+8cAB^3\,,
\eea
\be\label{8.31}
(1-m)G^2(AF-BD)(AF+3BD)=g^2F^4-4B^2(aF^2+bBF+cB^2)\,.
\ee
In the special case of $m=1$, the solution (\ref{8.26}) goes
over to the hyperbolic soliton solution (\ref{8.18}).

In the special case of $D=0$, $F=1$, we obtain a simpler solution
\be\label{8.32}
\phi=\sqrt{A\dn(Gx+x_0,m)+B}\,,
\ee
provided
\be\label{8.33}
c>0\,,~~G^2=\frac{9b^2-32ac}{8c(2-m)}\,,~~A^2=\frac{3G^2}{4c}\,,
B=-\frac{3b}{8c}\,,
\ee
while $g$ for this solution is given by
\bea\label{8.34}
&&g^2=4B^2(a+bB+cB^2)+(1-m) A^2 G^2 \nonumber \\
&&=\frac{3}{[32(2-m)]^2 c^3}[192b^2 ac(m^2+8m-8) \nonumber \\
&&+(64ac)^2 (1-m)+9b^4(16-16m-5m^2)]\,.
\eea

Like the $\cn$ case, in this case too solutions exist in case $A=0$
(and assuming $B=1$ without any loss of generality). In particular,
\be\label{8.35}
\phi=\frac{1}{\sqrt{D\dn(Gx+x_0,m)+F}}\,,
\ee
is an exact solution to Eq. (\ref{5a}) provided
\bea\label{8.36}
&&G^2=g^2 D^2\,,~~3(1-m)D^2 G^2=-g^2 F^4+4(aF^2+bF+c)\,, \nonumber \\
&&(2-m)G^2 =2(3g^2 F^2-2a)\,,~~4g^2 F^3=b+4aF\,.
\eea
Note that this solution is valid for all values of $m$ including the
hyperbolic limit of $m=1$.

However, unlike the $\cn$ case, in this case solution exists even in case
$B=0$ (and assuming $A=1$ without any loss of generality). In particular
\be\label{8.37}
\phi=\sqrt{\frac{\dn(Gx+x_0,m)}{D\dn(Gx+x_0,m)+F}}\,,
\ee
is an exact solution to Eq. (\ref{5a}) provided
\bea\label{8.38}
&&(1-m)G^2=g^2 F^2\,,~~3 F^2 G^2=-g^2 D^4+4(aD^2+bD+c)\,, \nonumber \\
&&(2-m)G^2 =2(3g^2 D^2-2a)\,,~~4g^2 D^3=b+4aD\,.
\eea
Note that such a solution does not exist in the hyperbolic limit
so long as $g \ne 0$.

{\bf Solution IV}

It is easily shown that
\be\label{8.39}
\phi=\sqrt{\frac{A\sn^2(Gx+x_0,m)+B}{D\sn^2(Gx+x_0,m)+F}}\,,
\ee
is an exact solution to the field Eq. (\ref{5a}) provided the following five
coupled equations are satisfied:
\be\label{8.40}
4mG^2 AD(AF-BD)=-g^2 D^4+4A^2(aD^2+bAD+cA^2)\,,
\ee
\bea\label{8.41}
&&G^2(AF-BD)[2mAF+2(1+m)AD]=g^2 D^3 F-2aAD(AF+BD) \nonumber \\
&&-bA^2(AF+3BD)-4cA^3 B\,,
\eea
\bea\label{8.42}
&&2G^2(AF-BD)[3mBF-(1+m)AF+(1+m)BD-3AD] \nonumber \\
&&=3g^2 D^2 F^2 -2a(A^2 F^2+B^2 D^2+4ABDF) \nonumber \\
&&-6bAB(AF+BD)-12cA^2 B^2\,,
\eea
\bea\label{8.43}
&&2G^2(AF-BD)B[D+(1+m)F]=-g^2 DF^3+2aBF(AF+BD) \nonumber \\
&&+bB^2(3AF+BD)+4cAB^3\,,
\eea
\be\label{8.44}
4G^2(AF-BD)BF=g^2F^4-4B^2(aF^2+bBF+cB^2)\,.
\ee
Thus we have five coupled equations relating the five parameters $A,B,D,G,g$.
In the special case of $m=1$, the solution (\ref{8.39}) goes
over to the hyperbolic soliton solution
\be\label{8.45}
\phi=\sqrt{\frac{A\tanh^2 (Gx+x_0)+B}{D\tanh^2 (Gx+x_0)+F}}\,,
\ee
while in the limit $m=0$ it goes over to the trigonometric solution
\be\label{8.46}
\phi=\sqrt{\frac{A\sin^2 (Gx+x_0)+B}{D\sin^2 (Gx+x_0)+F}}\,.
\ee

{\bf Solution V}

The field Eq. (\ref{5a}) also has a remarkable nonperiodic solution given
by
\be\label{8.47}
\phi=\sqrt{\frac{Ax^2+B}{Dx^2+F}}\,,
\ee
provided the following five relations are satisfied
\be\label{8.48}
g^2 D^4 =4A^2(aD^2+bAD+cA^2)\,,
\ee
\be\label{8.49}
g^2 D^3 F=2aAD(AF+BD)+bA^2(AF+3BD)+4cA^3 B\,,
\ee
\bea\label{8.50}
&&6AD(AF-BD)+3g^2 D^2 F^2=2a(A^2 F^2+B^2 D^2+4ABDF) \nonumber \\
&&+6bAB(AF+BD)+12cA^2 B^2\,,
\eea
\bea\label{8.51}
&&2BD(AF-BD)+g^2 DF^3=2aBF(AF+BD) \nonumber \\
&&+bB^2(3AF+BD)+4cAB^3\,,
\eea
\be\label{8.52}
g^2F^4=4BD(AF-BD)+4B^2(aF^2+bBF+cB^2)\,.
\ee


\subsection{Stability of nonlinear-phase-modulated solutions}

We now briefly discuss the stability of the nonlinear-phase-modulated solutions presented in this 
section.  We emphasize that dynamical instability is unaffected by a non-trivial linear phase factor, corresponding to a traveling frame $x \rightarrow x - v t$ as described in Sec. \ref{intro}. However, 
it is not clear {\it a priori} what the physical relevance may be of a solution with nonlinear phase, 
such as thosepresented here. Also, none of the solutions presented here are related in any way 
to the stable family of solutions from the preceding section.  In fact, only Solution I could be 
considered among the dark soliton family, but it does not  exist for $m=1$, and is not similar in 
form to the stable solution from the previous section. Additionally, the question of finding these 
solutions numerically is significantly more challenging than the trivial phase solutions, at least for 
the periodic amplitude ones, owing to the necessity of finding suitable parameters
which satisfy systems of coupled nonlinear equations and lead to admissible (real-valued) and 
nontrivial solutions.  We do not expect these solutions to be more stable than their trivial phase 
counterparts from the previous section. Therefore, while the solutions presented herein are valid 
and interesting in their own right, we do not investigate their stability in detail here.  Nonetheless, 
for illustrative purposes we briefly explain the methodology and then present two solutions and 
their stability, one at the hyperbolic limit, $m=1$, and one elliptic solution.

The criteria in this section for the 10 parameters involved in the equation and corresponding solution 
consist, in general, of 5 coupled nonlinear equations, for which one can choose 5 parameters and 
solve for the remaining 5 with Newton's method, for instance.  In this way, one often finds that 
A/B=D/F, which is a trivial solution, or that the parameters either are imaginary or lead to imaginary 
solutions.

Once the solution $\phi$ to Eq. \ref{5a} is found, the solution $u=\phi \exp{\{i g/2 \eta\}}$ must be 
found in a numerical domain in order to examine its stability.  This leads to the additional resonance 
condition that there exist integers $n$ and $k$ such that

\be
n \pi/ g = k P(m,G)
\ee

\noindent where $P(m,G) = 4 K(m)/G$ is the period of the elliptic function $\phi$ and $K(m)$ is the 
complete elliptic integral of the first kind.  The domain has to be then truncated at the period of $\eta$, 
and of $\phi$ in case $m \neq 1$.  If $m=1$, we choose one period of $\eta$ as the domain, i.e. 
$\{L; \eta = \pi/g\}$.  For $m \neq 1$, we determine $(n^*,k^*)={\rm argmin}\{(n,k); n \pi g = k P(m,G)\}$ 
and then truncate the domain at $\{L;  \eta = n \pi/g\}$.

\begin{figure}
\includegraphics[width=70mm]{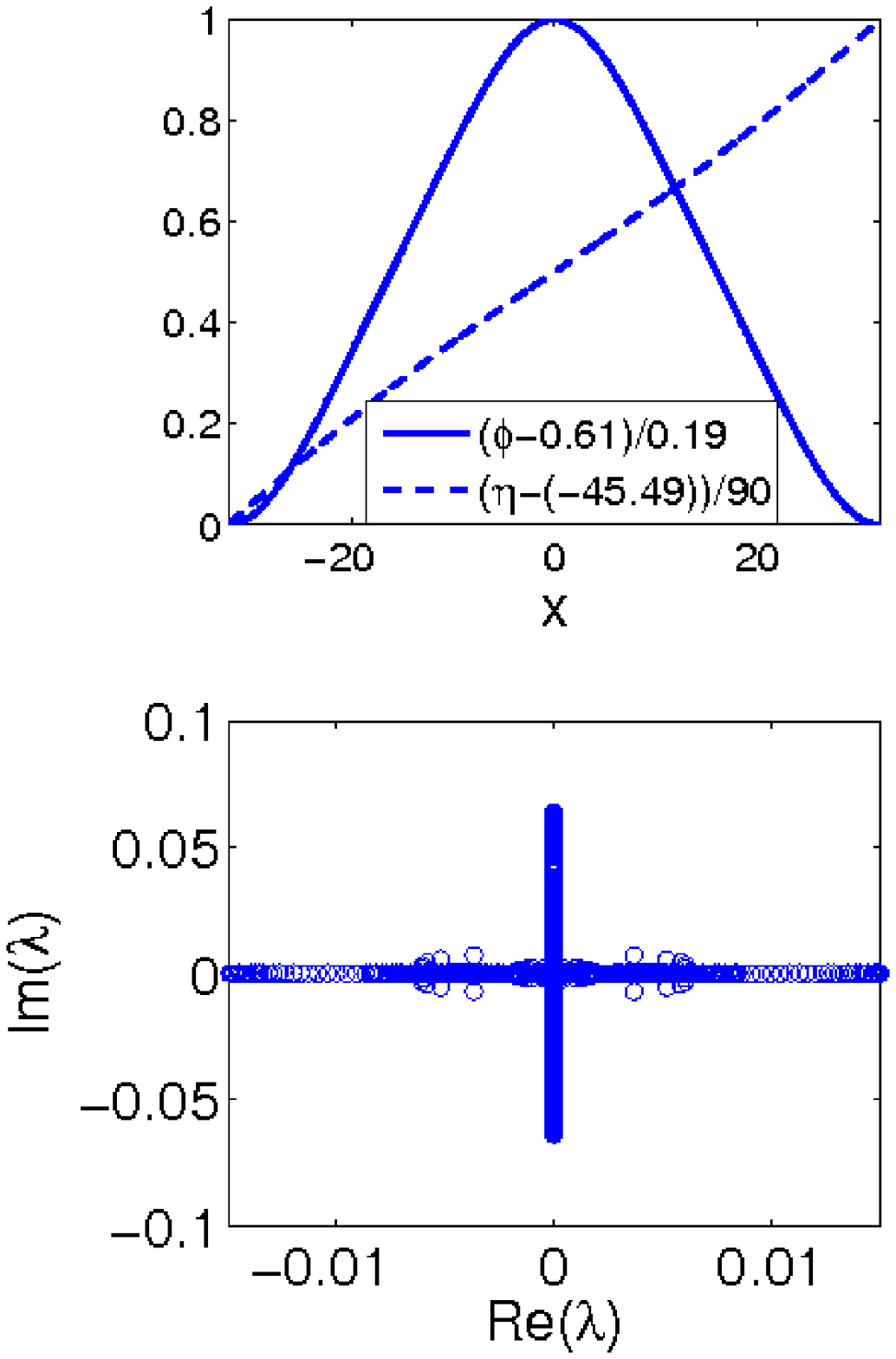}
\includegraphics[width=70mm]{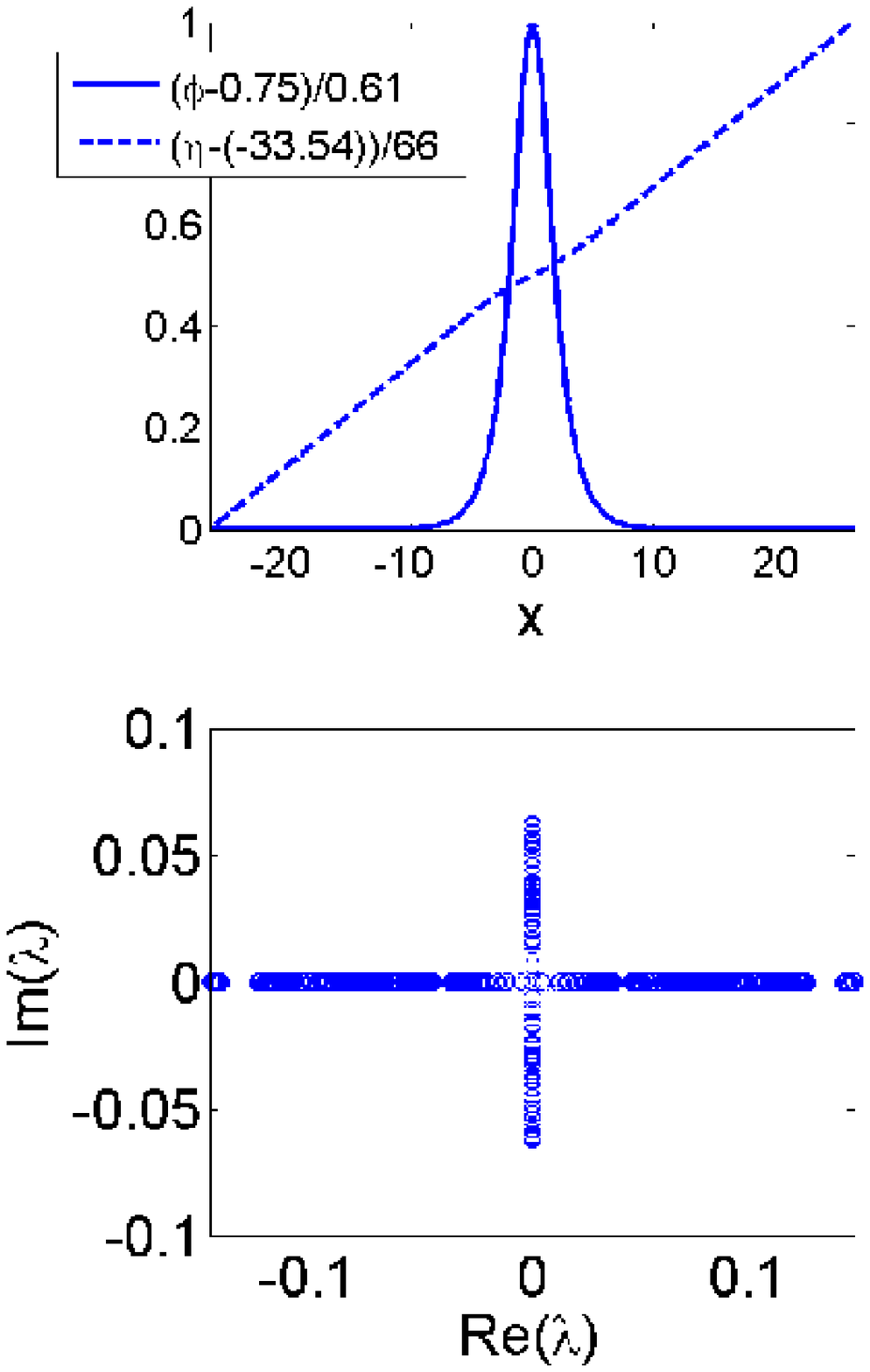}
\caption{(Color Online) The top panels depict solution and phase and the bottom panels show the linear spectra.  The left-hand solution is the elliptic solution given in Eq. (\ref{8.20}) with $(a,b,c,m,g,A,B,D,F,G)= (0.5077,-0.9501,0.5,0.5,0.4145,0.1,0.7126,0,1,0.1155)$, while the
right-hand solution is for the hyperbolic limit with $(a,b,c,m,g,A,B,D,F,G)= 
(1,-2,1,1,0.375,0.6124,0.75,0,1,0.7071)$.}
\label{nlp_sols}
\end{figure}

As displayed in Fig. \ref{nlp_sols}, the stability does not differ significantly qualitatively from the solutions with trivial phase (compare with the solutions in Figs. \ref{F5},\ref{F8} top), except the hyperbolic solution is now periodic in phase, and, hence, has a continuum of unstable eigendirections.  The dynamics of both these solutions, when perturbed with additive noise, results in collapse, as in the case of trivial phase.

\section{Conclusions}
\label{conclusion}

To conclude, a large array of exact analytical solutions to
various field theories were found
leading directly to various exact solutions of cubic-quintic
NLS and their stability and dynamics were studied
numerically.
The majority of these are oscillatory solutions which are unstable,
however, some are found to be linearly stable, such as some elliptic
solutions from the
dark-soliton family solution II for appropriate parameter regimes.
Also many hyperbolic dark-solitons were found
to be stable and quite robust over a range of parameter values.  The
hyperbolic bright soliton solutions were found to be unstable to
collapse when the cubic
term is defocusing or nonexistent and the quintic term is focusing,
while they were
found to be always stable when the cubic term is focusing, for both focusing
and defocusing quintic term (at least within the parametric regions
examined herein).  Some periodic solutions with
defocusing and focusing cubic and quintic nonlinear terms, respectively,
also displayed collapse phenomenon, while ones with a focusing cubic term
did not.

\begin{appendix}
\section{Solutions of $\phi^2$-$\phi^3$-$\phi^4$ and hence Quadratic-Cubic NLS
 Model}
\label{A1}

Let us consider solutions of field Eq. (\ref{7}). As explained above, once
these solutions are obtained, then the solution of the quadratic-cubic
 equation are
immediately obtained from here by using Eqs. (\ref{1}), (\ref{2}) and
(\ref{6}). We list below eight distinct periodic solutions to the field
Eq. (\ref{7}). In each case, we also mention the values of the parameters
$a,b,c$, in particular, if they are positive or negative.
However, instead of Eq. (\ref{7}) we use slightly different form given by
\be\label{3.1}
\phi_{xx}+2a\phi+3b\phi^2+4c\phi^3=0\,.
\ee
This is done so that one can easily pick up the solutions recently obtained by
us in a related paper on coupled $\phi^2$-$\phi^3$-$\phi^4$ field theory \cite{ks2}.

{\bf Solution I}

It is easily shown \cite{ks2} that
\be\label{3.2}
\phi=F+A\sn(Bx+x_0,m)\,,
\ee
is an exact solution to the field Eq. (\ref{3.1}) provided
\be\label{3.3}
(1+m) B^2=-a\,,~~2(1+m)cA^2=ma\,,~~b=-4cF\,,~~b^2=4ac\,.
\ee
Thus this solution is valid provided $a<0,c<0$ while $b$ could be
positive or negative.

In the limit $m=1$ the periodic solution (\ref{3.2}) goes over to
 the dark soliton solution
\be\label{3.4}
\phi=F+A\tanh(Bx+x_0)\,,
\ee
provided
\be\label{3.5}
2B^2=-a\,,~~4cA^2=a\,,~~b=-4cF\,,~~b^2=4ac\,.
\ee
Thus the dark soliton solution exists to field Eq. (\ref{3.1}) provided
$a<0,c<0$ while $b$ could be positive or negative.

{\bf Solution II}

It is easily shown \cite{ks2} that
\be\label{3.6}
\phi=F+A\cn(Bx+x_0,m)\,,
\ee
is an exact solution to the field Eq. (\ref{3.1}) provided
\be\label{3.7}
(2m-1) B^2=a\,,~~2cA^2=mB^2\,,~~b^2=4ac\,,b=-4cF\,.
\ee
Thus this solution is valid provided $a>0,c>0$ while $b$ could be positive or
negative. Note also that this solution exists only if $m > 1/2$.

In the limit $m=1$ the periodic solution (\ref{3.6}) goes over to
 the bright soliton solution
\be\label{3.8}
\phi=F+A\sech(Bx+x_0)\,,
\ee
provided
\be\label{3.9}
B^2=a\,,~~2cA^2=B^2\,,~~b^2=4ac\,,~~b=-4cF\,.
\ee
Thus the bright soliton solution exists to field Eq. (\ref{3.1}) provided
$a>0,c>0$ while $b$ could be either positive or negative.

{\bf Solution III}

It is easily shown \cite{ks2} that
\be\label{3.10}
\phi=F+A\dn(Bx+x_0,m)\,,
\ee
is an exact solution to the field Eq. (\ref{3.1}) provided
\be\label{3.11}
(2m-1) B^2=a\,,~~2cA^2=B^2\,,~~b^2=4ac\,,~~b=-4cF\,.
\ee
Thus this solution is valid provided $a>0,c>0$ while b could be either
positive or negative. Note also that unlike solution II, this solution
is valid for any $m (0 \le m \le 1$).

In the limit $m=1$ the periodic solution (\ref{3.10}) goes over to
 the bright soliton solution (\ref{3.8}) and hence satisfy constraints
given by Eq. (\ref{3.9}).

{\bf Solution IV}

It is easily shown \cite{ks2} that
\be\label{3.12}
\phi=F+\frac{A\sn(Dx+x_0,m)}{1+B\dn(Dx+x_0,m)}\,,
\ee
is an exact solution to the field Eq. (\ref{3.1}) provided
\be\label{3.13}
B=1\,,~~(2-m) D^2=-2a\,,~~4(2-m)cA^2=m^2a\,,~~b^2=4ac\,,~~b=-4cF\,.
\ee
Thus this solution is valid provided $a<0,c<0$ while $b$ could be either
positive or negative.

In the limit $m=1$ the periodic solution (\ref{3.12}) goes over to
 the dark soliton solution
\be\label{3.14}
\phi=F+\frac{A\tanh(Dx+x_0)}{1+B\sech(Dx+x_0)}\,,
\ee
provided
\be\label{3.15}
B=1\,,~~D^2=-2a\,,~~4cA^2=a\,,~~b^2=4ac\,,~~b=-4cF\,.
\ee
Thus the dark soliton solution exists to field Eq. (\ref{3.1}) provided
$a<0,c<0$ while $b$ could have either sign.

{\bf Solution V}

It is easily shown \cite{ks2} that
\be\label{3.16}
\phi=F+\frac{A\cn(Dx+x_0,m)}{\sqrt{1-m}+B\dn(Dx+x_0,m)}\,,
\ee
is an exact solution to the field Eq. (\ref{3.1}) provided
the parameters satisfy the same constraints as given by Eq. (\ref{3.13}).
Thus this solution is valid provided $a<0,c<0$ while $b$ could have
either sign.

In the limit $m=1$ the periodic solution (\ref{3.16}) goes over to
a constant, i.e.
\be\label{3.18}
\phi=F+A\,,
\ee
where $F,A$ are as given by Eq. (\ref{3.15}).

{\bf Solution IV}

It is easily shown \cite{ks2} that
\be\label{3.19}
\phi=F+\frac{A\tanh(Dx+x_0)}{1+B\tanh(Dx+x_0)}\,,
\ee
is an exact solution to the field Eq. (\ref{3.1}) provided
\be\label{3.20}
~~2D^2=-a\,,~~A=F(1-B)\,,~~b^2=4ac\,,~~Fb=-(1+B)a\,.
\ee
Thus this solution is valid provided $a<0,c<0$ while $b$ could be either
positive or negative.

{\bf Solution VII}

It is easily shown \cite{ks2} that
\be\label{3.21}
\phi=\frac{A\sech(Dx+x_0)}{1+B\sech(Dx+x_0)}\,,
\ee
is an exact solution to the field Eq. (\ref{3.1}) provided
\be\label{3.22}
A=\frac{-2a}{\sqrt{b^2-4ac}}\,,~~D^2=-2a\,,~~B=\frac{b}{\sqrt{b^2-4ac}}\,.
\ee
Thus this solution is valid provided $a<0$ while $b,c$ could have either
sign. In particular, if $c>0$ then $b$ can have either sign while if $c<0$
then $b>0$. This solution can also be rewritten as
\be
\phi=\frac{-2a}{b+\sqrt{b^2-4ac}\cosh[\sqrt{2a}(x+x_0)]}\,.
\ee

{\bf Solution VIII}

It is easily shown \cite{ks2} that
\be\label{3.23}
\phi=F+\frac{A\sech(Dx+x_0)}{1+B\sech(Dx+x_0)}\,,
\ee
is also an exact solution to the field Eq. (\ref{3.1}) provided
\be\label{3.24}
A=-\frac{4a+3bF}{\sqrt{b^2+2bcF}}\,,~~D^2=3bF+4a\,,~~
B=-\frac{b+4cF}{\sqrt{b^2+2bcF}}\,.
\ee
Thus this solution is valid provided $b>0,c<0$ while $a>(<$ or $=) 0$ depending
on if $m< (>$ or $=) 1/5$.

{\bf Solution IX}

It is easily shown that
\be\label{3.25}
\phi=\frac{Ax^2+B}{Dx^2+E}\,,
\ee
is an exact solution to the field Eq. (\ref{3.1}) provided
\be\label{3.26}
AE=-3BD\,,~~\frac{A}{B}=-\frac{4a}{3}\,,~~9b^2=32ac\,,~~
\frac{A}{D}=-\frac{4a}{3b}\,.
\ee
Thus this solution is only valid if $a>0,c>0$ while $b$ could be either
positive or negative.

\section{Solutions of $\phi^2$-$\phi^4$ and hence Standard (Cubic) NLS
 Model}
 \label{A2}

Finally, for completeness, we consider solutions of field Eq. (\ref{7}).
in case $b=0$. As explained above, once
these solutions are obtained, then the solution of the standard (cubic)
NLS equation are immediately obtained from here by using Eqs. (\ref{1}),
(\ref{2}) and
(\ref{6}). We list below three distinct periodic solutions to the field
Eq. (\ref{7}) with $b=0$. In each case, we also mention the values of the
parameters $a$ and $c$, in particular, if they are positive or negative.
However, instead of Eq. (\ref{7}) (with $b=0$)
we use slightly different form given by
\be\label{4.1}
\phi_{xx}+2a\phi+4c\phi^3=0\,.
\ee
This is done so that one can easily pick up the solutions recently obtained by
us in a related paper on coupled $\phi^2-\phi^4$ field theory \cite{ks3}.

{\bf Solution I}

It is easily shown \cite{ks3} that
\be\label{4.2}
\phi=A\sn(Bx+x_0,m)\,,
\ee
is an exact solution to the field Eq. (\ref{4.1}) provided
\be\label{4.3}
(1+m) B^2=2a\,,~~(1+m)cA^2=-ma\,.
\ee
Thus this solution is valid provided $a>0,c<0$.

In the limit $m=1$ the periodic solution (\ref{4.2}) goes over to
 the dark soliton solution
\be\label{4.4}
\phi=A\tanh(Bx+x_0)\,,
\ee
provided
\be\label{4.5}
B^2=a\,,~~2cA^2=-a\,.
\ee
Thus the dark soliton solution exists to field Eq. (\ref{4.1}) provided
$a>0,c<0$.

{\bf Solution II}

It is easily shown \cite{ks3} that
\be\label{4.6}
\phi=A\cn(Bx+x_0,m)\,,
\ee
is an exact solution to the field Eq. (\ref{4.1}) provided
\be\label{4.7}
(2m-1) B^2=-2a\,,~~(2m-1)cA^2=-ma\,.
\ee
Thus this solution is valid provided $c>0$ while $a < (>$ or $=) 0$ depending
on if $m > (<$ or $=) 1/2$.

In the limit $m=1$ the periodic solution (\ref{4.6}) goes over to
 the bright soliton solution
\be\label{4.8}
\phi=A\sech(Bx+x_0)\,,
\ee
provided
\be\label{4.9}
B^2=-2a\,,~~cA^2=-a\,.
\ee
Thus the bright soliton solution exists to field Eq. (\ref{4.1}) provided
$a<0,c>0$.

{\bf Solution III}

It is easily shown \cite{ks3} that
\be\label{4.10}
\phi=A\dn(Bx+x_0,m)\,,
\ee
is an exact solution to the field Eq. (\ref{4.1}) provided
\be\label{4.11}
(2-m) B^2=-2a\,,~~(2-m)cA^2=-a\,.
\ee
Thus this solution is valid provided $a<0,c>0$.

In the limit $m=1$ the periodic solution (\ref{4.10}) goes over to
 the bright soliton solution (\ref{4.8}) and hence satisfy constraints
given by Eq. (\ref{4.9}).

\end{appendix}

\end{document}